\documentclass[preprint,3p,12pt]{elsarticle}
\usepackage{titlesec}
\titleformat{\paragraph}[runin]{\normalfont\normalsize\bfseries}{\theparagraph}{1em}{}

\usepackage[utf8]{inputenc}
\usepackage[T1]{fontenc}
\usepackage{calc}
\usepackage{subfigure}
\usepackage{titlesec}
\usepackage{graphicx}
\usepackage{mathtools}
\usepackage{xcolor}
\usepackage{multirow}
\usepackage{comment}
\usepackage{ulem}
\usepackage{booktabs}
\usepackage{chemformula}
\usepackage{soul}
\usepackage{longtable}

\usepackage[linesnumbered,ruled,vlined]{algorithm2e}
\usepackage{amssymb}

\journal{Elsevier}
\SetKwInput{KwInput}{Input}                
\SetKwInput{KwOutput}{Output}              

\graphicspath{{img/}}

\begin{document}

\begin{frontmatter}


  \title{Process-structure-property relation for elastoplastic behavior of soft nanocomposites with agglomerates and interfacial gradient}


  \author[label1]{Prajakta Prabhune}
  \author[label1]{Anlan Chen}
  \author[label2]{Yigitcan Comlek}
  \author[label2]{Wei Chen}
  \author[label1]{L. Catherine Brinson\corref{cor1}}
    \cortext[cor1]{Corresponding author}
  \ead{cate.brinson@duke.edu}
  \address[label1]{Thomas Lord Department of Mechanical Engineering and Materials Science, Duke University, Durham, USA}
 \address[label2]{Department of Mechanical Engineering, Northwestern University, Evanston, USA}


\begin{abstract}

Polymer nanocomposites, inherently tailorable materials, are potentially capable of providing higher strength to weight ratio than conventional hard metals. However, their disordered nature makes processing control and hence tailoring properties to desired target values a challenge. Additionally, the interfacial region, also called the interphase, is a critical material phase in these heterogeneous materials and its extent depends on variety of microstructure features like particle loading and dispersion or inter-particle distances. 
Understanding process-structure-property (PSP) relation can provide guidelines for process and constituents’ design. Our work explores nuances of PSP relation for polymer nanocomposites with attractive pairing between particles and the bulk polymer. Past works have shown that particle functionalization can help tweak these interactions in attractive or repulsive type and can cause slow or fast decay of stiffness properties in polymer nanocomposites. In this work, we develop a material model that can represent decay for small strain elastoplastic(Young’s modulus and yield strength) properties in interfacial regions and simulate representative or statistical volume element behavior.  The interfacial elastoplastic material model is devised by combining local stiffness and glass transition measurements from atomic force microscopy and fluorescence microscopy. 
This model is combined with a microstructural design of experiments for agglomerated nanocomposite systems. Agglomerations are particle aggregations arising from processing artifacts. Twin screw extrusion process can reduce extent of aggregation in hot pressed samples via erosion or rupture depending on screw rpms and torque. We connect this process-structure relation to structure-property relation that emerges from our study. We discover that balancing between local stress concentration zones (SCZ) and interfacial property decay governs how fast yield stress can improve by breaking down agglomeration via erosion. Rupture is relatively less effective in helping improve nanocomposite yield strength. We also observe an inflection point where incremental increase brought on by rupture is slowed due to increasing SCZ and saturation in interphase percolation. 

\end{abstract}


\end{frontmatter}


\newcommand{\revision}[2]{\sout{#1} \textcolor{red}{(#2)}}

\section{Introduction} \label{Introduction}
Understanding the process-structure-property (PSP) relation is fundamental to design and manufacturing of advanced heterogeneous and disordered materials such as polymer nanocomposites that are deployed in variety of technology domains. For the ease of their handling and high strength to weight ratio, 3D printing processes are increasingly using polymer nanocomposites for fabricating precision components used in aerospace, electrical, electronics, electrochemical and nanomedicine systems \cite{al2021additive,natarajan2022processing,shameem2021brief}. Characterising elastoplastic material properties of nanocomposites is critical from the perspective of mechanical design of parts and components critical to structural integrity and load bearing purposes \cite{naskar2016polymer}.

However, establishing PSP relation for elasto-plastic behavior of nanocomposites is especially challenging due to lack of process control for target microstructure and/or properties, difficulties encountered in characterising 3D microstructures accurately due to the stochastic nature of material microstructure \cite{bostanabad2018computational}, lack of nanomechancial property measurements beyond elastic limits \cite{saito2024pushing,collinson2021best}, and time and effort involved in systematic experimentation to characterize bulk properties of an extensive number of samples\cite{gianola2023advances,sheridan2024botts}. Combined methods of computational mechanics with methods of artificial reconstruction of statistically equivalent microstructure can be used to overcome these challenges and develop multiscale modelling and design tools \cite{prabhune2023design}.

An especially important and poorly understood phenomenon is the impact of particle dispersion and aggregation on nanocomposite properties. If particle-particle interaction is relatively attractive compared to polymer-particle interaction, they tend to form larger, up to microsized particle aggregates called agglomerations during nanoparticle syntheis as well as nanocomposite manufacturing process \cite{dorigato2010linear, oberdisse2006aggregation, li2007nucleation, cassagnau2008melt}. Agglomerations can be seen as particle dispersion anomalies in nanocomposites that result due to lack of processing control. Past research has witnessed a transition from brittle to ductile behaviour in nanocomposite mechanical properties that accompanies transition in particle aggregate size from a fraction of micrometer to a few nanometers \cite{matsushige1976pressure,wee2024brittle,ash2004mechanical}. Overall, the consensus is that high particle loading, which usually is prone to presence of agglomerations, tends to degrade mechanical properties. Additionally the size effect studies indicate that nanosized inclusions are more effective at property improvement than microsizes inclusions \cite{cho2006effect,hamming2009effects}. Multiple approaches have been attempted to explain these observed property trends. 

In top-down approaches, micromechanics theories are modified to account for the presence of  microstructural anomalies such as aggregates and agglomerations; interfacial effects such as interfacial adhesion/debonding and interphase properties \cite{fornes2003modeling,wang2004prediction,fisher2003fiber,hbaieb2007modelling,liu2008reinforcing,luo2003characterization,pukanszky1990influence,pukanszky1993mechanism,zare2016study}. These modifications are in the form of model or material parameters which are deduced by fitting the model against available test data. Often, such approaches are limited by their inability to capture local effects like local property decay in teh interphase region, local variation in particle dispersion and volume fraction and their effect on interphase percolation. For this very reason, these studies have been able to predict linear properties very well but not nonlinear ones. These models show a significant gap between predictions and experiments for nonlinear properties such as yield and fracture since local effects are more prominent in determining effective yielding and failure. Additionally, most of the parameters in such models are empirical with limited physical meaning. Thus such parameters are not useful for developing any mechanistic insight into material behavior. 

In contrast, bottom-up approaches try to combine phenomenological micro mechanical knowledge with nanoscale measurements and numerical simulations to inform modelling through scales \cite{peng2012modeling,odegard2017modeling,LI2019100277}. Local changes in polymer properties near particles are usually represented by an intrinsic material phase called the interphase with modified material properties derived from base polymer properties. Experimental characterization efforts at nano and microscale have established that interfacial interactions and confinement causes altered polymer chain mobility in the vicinity and resulting in altered physical material properties such as glass transition temperature, elastic modulus, and dielectric constant. Several attempts have been made to experimentally characterize local property variations at both nano and microscale \cite{rittigstein2007model,ellison2003distribution,zhang_stiffness_2017,mapesa2020wetting}. Atomic Force Microscopy (AFM) characterization of the local modulus on sandwich model composites has established a gradual property decay away from the interface between matrix and particle, up to a few hundred nanometers, and non-trivial compounding in the interphase region \cite{zhang_stiffness_2017,zhang_determination_2018}. Li et al. incorporated this understanding in a 2D interphase model for dispersed nanocomposite systems \cite{LI2019100277}. This model assumes gradient interphase for each particle with multibody compounding effect to simulate elastic and frequency-based viscoelastic properties of nanocomposite samples.
modeled by equations below \cite{LI2019100277}

To accurately simulate elastoplastic material behavior of polymer nanocomposites, the essential features of stress-strain relation for each material phase involved are the Young's modulus and yield stress. Local nanoscale measurements of such data in situ in real nanocomposites are limited to a few measurements of elastic properties at room temperature around isolated particles \cite{collinson2019deconvolution} because of the many challenges associated with sample preparation and small scale experimentation. Previous studies on characterizing the local mechanical behaviors are limited and/or are performed on pure polymers only \cite{du2001study,kolluru2018afm,min2019}. To the best of the authors' knowledge, there is no direct data available on local non-linear mechanical property variations in nanocomposite material system. Hence, in this work, we combine our understanding of local elastic modulus data, the effect of interfacial interactions on molecular mobility, and the glass transition temperature on the macroscopic non-linear mechanical behavior of amorphous polymers to model the inelastic behavior of the interphase. 

We employ this elastoplastic interphase material model to simulate a design of experiments(DoE) devised to elucidate PSP relation for naocomposites with consideration to various dispersion levels due to agglomeration anomalies and 3 categories of interfacial gradient schemes, specifically coupled slow decay gradient, coupled fast decay gradient and decoupled gradient. We form an understanding of how microstructure parameters such as particle volume fraction, volume fraction forming agglomerations and number of agglomerations is related to effective nanocomposite properties. We compare the effect of two processes, rupture and erosion, used to break down agglomeration to improve dispersion through tuning of processing parameters such as screw speed (rpm) and mixing energy etc. We also study the mechanism of yielding process in polymer nanocomposites with presence of a graded interphase. In this process, we devised a measure of local yield resistance which can be used as predictor of local yield progression pattern that we call yield percolation network. Essentially, yield resistance maps calculated using linear stress concentration field values and local interphase yield strength act as a blueprint of yield percolation network. 

\section{Methods} \label{methods}
In this section, we first describe (\ref{percolation rules}) how interfacial gradient percolation is modeled using a combination of single body property decay and compounding of multi-body decay within dispersed nanocomposites. Next, we describe how an elastoplastic interphase model is built \ref{El_Pl_Interphase}. Then we explain the method we use to generate artificial microstructures that are statistically equivalent and can serve a purpose of being a statistical volume element (SVE) for the material system in consideration (ref{mcr}). Further, we describe DoE that consists of microstructural space and three interphase schemes. This DoE (approximately 2000 data points) is simulated and analyzed to tease out the PSP relation of Young's modulus and Yield strength of polymer nancomposites (\ref{DoE}).

\subsection{Interphase percolation} \label{percolation rules}
\begin{figure}
    \centering
    \subfigure[stiffness decay with $\beta = 0.03$]{\includegraphics[width=.48\linewidth]{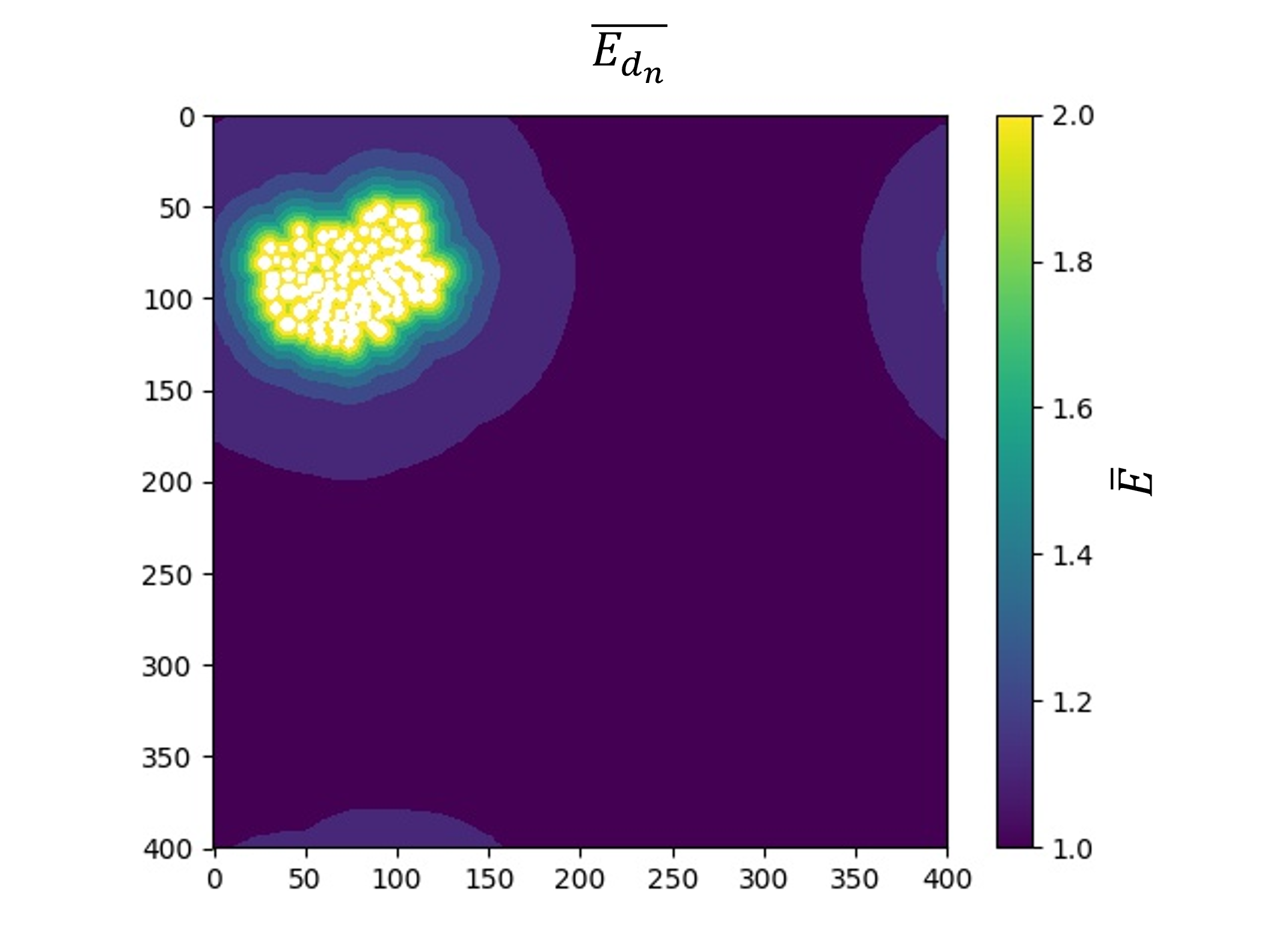}\label{E_singlebody}}
     \subfigure[$\Delta T_g$ decay with $\beta' = 0.03$]{\includegraphics[width=.48\linewidth]{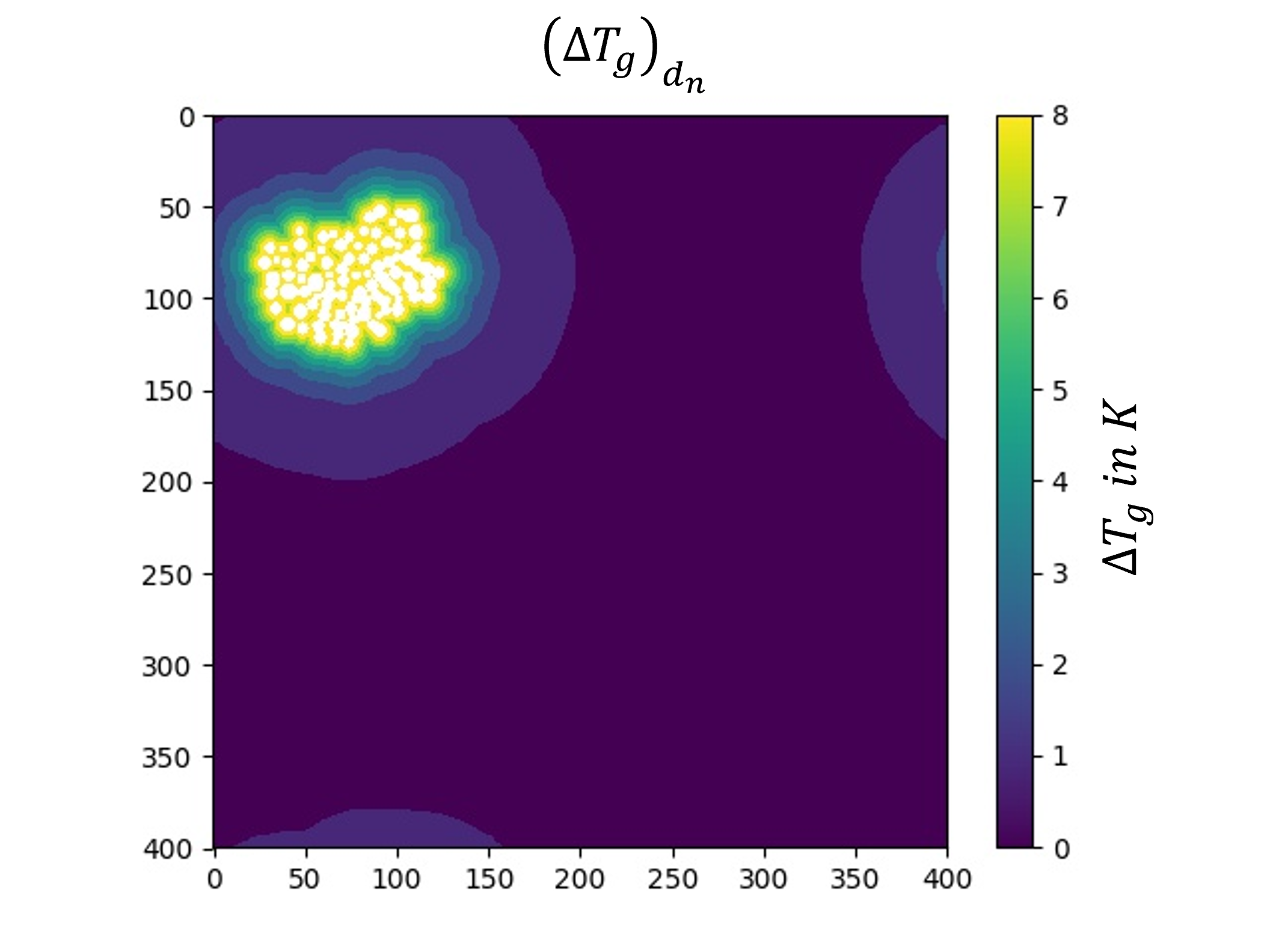}\label{deltaTg_singlebody}}
    \caption{Single body property decay demonstrated on one the SVEs with 2\% total VF that has single agglomeration with no isolated particle dispersion}
    \label{singlebody_decay}
\end{figure}

\begin{figure}
    \centering
    \subfigure[stiffness compounding effect for $\beta = 0.03$]{\includegraphics[width=.48\linewidth]{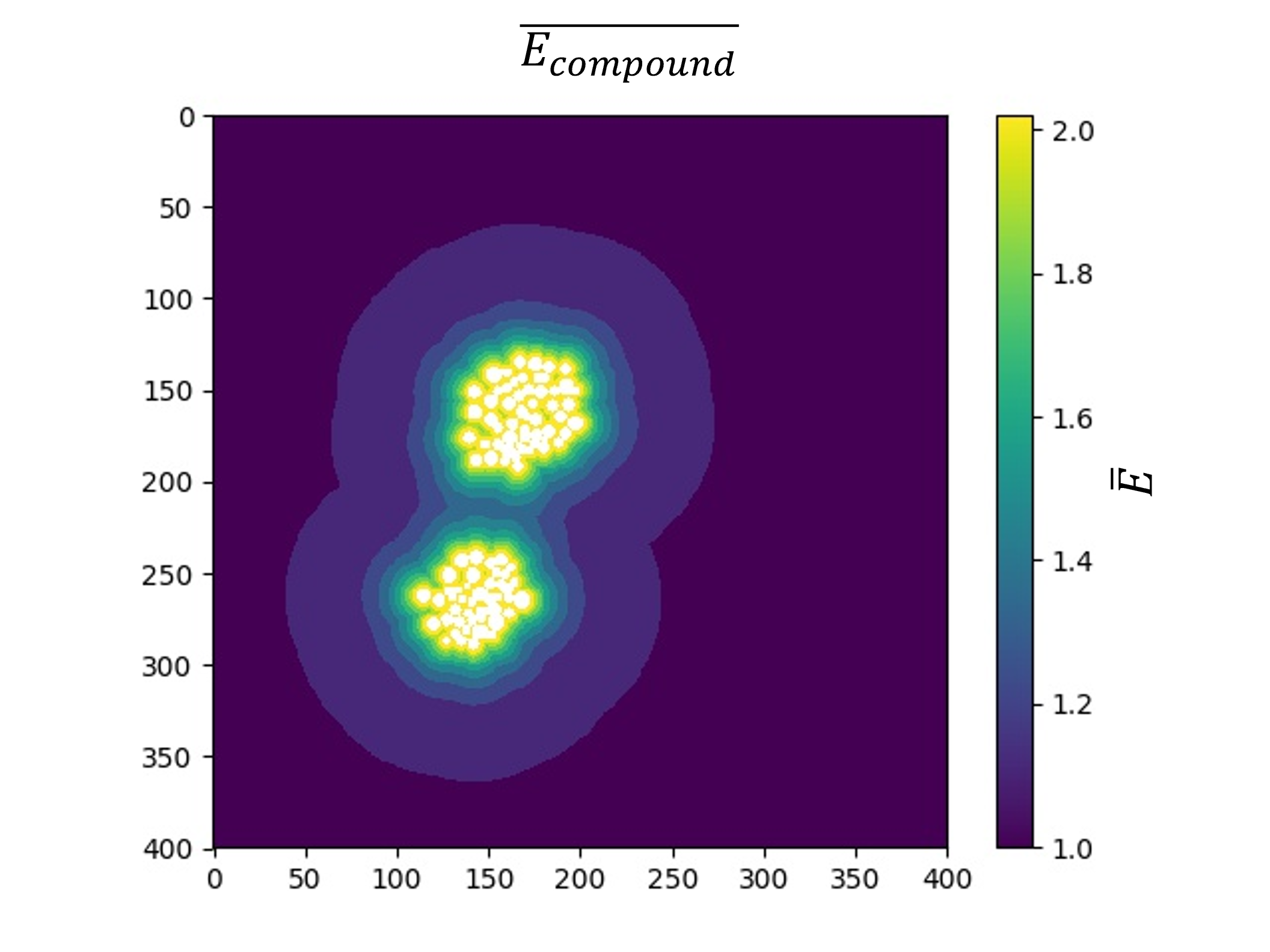}\label{E__compound2body}}
     \subfigure[$\Delta T_g$ compounding effect for $\beta' = 0.03$]{\includegraphics[width=.48\linewidth]{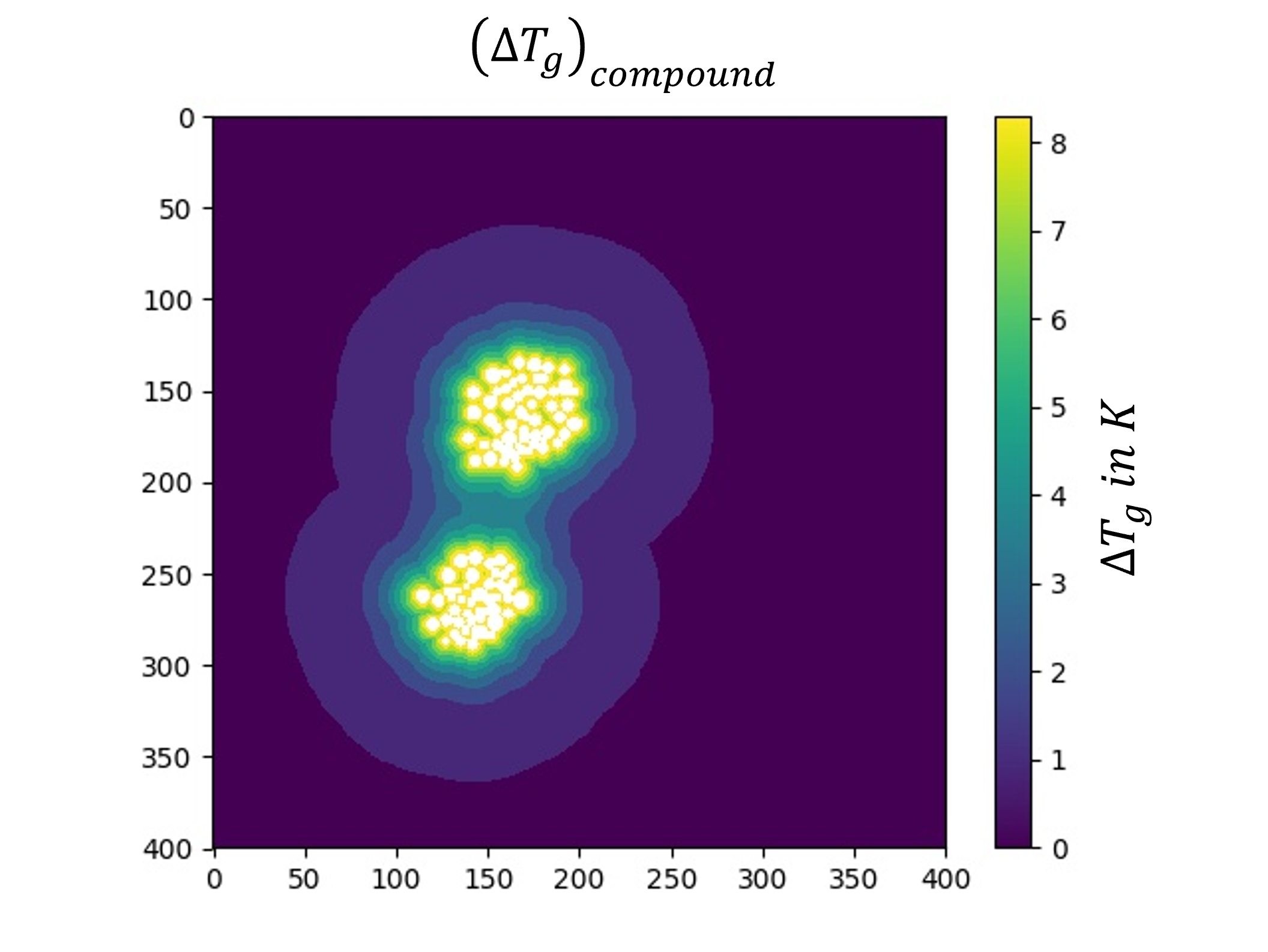}\label{deltaTg_compound2body}}
    \caption{Two body compounding demonstrated on one the SVEs with 5\% total VF that has two agglomerations with no isolated particle dispersion}
    \label{compounding}
\end{figure}

Here we use single body exponential decay and additive compounding \cite{LI2019100277}  to define changes within the interphase region. The expressions for Young's modulus $(E)$ and glass transition temperature $(\Delta T_g)$, are given by 

\begin{equation} \label{singleE}
    \overline{E_{d_n}} = \alpha e^{-{\beta d_n}}+1
\end{equation}
\begin{equation}\label{compoundE}
    \overline{E_{compound}} = \eta\left[\sum_{n=1}^{N}\left(\overline{E_{d_n}}-1\right)\right]+\zeta+1
\end{equation}

\begin{equation} \label{single_delTg}
    (\Delta T_g)_{d_n} = \alpha^{'}e^{-{\beta^{'}}d_n}+1
\end{equation}
\begin{equation}\label{compound_delTg}
    (\Delta T_g)_{compound} = \eta^{'}\left[\sum_{n=1}^{N}\left((\Delta T_g)_{d_n}-1\right)\right]+\zeta^{'}+1
\end{equation}
Here, Equation \ref{singleE} computes gradient decay of properties with respect to interface of a single isolated particle or a single agglomeration. Equation \ref{compoundE} additively compounds effects of two or more isolated particles and/or agglomerations. $\overline{E_{d_n}}$ and $\overline{E_{compound}}$ are normalized moduli as a function of distance $d_n$ from the interface of aggregate number $n$. Similarly, Equation \ref{single_delTg} and \ref{compound_delTg} calculate change in local glass transition temperature based on single body effect and compounding. Parameters sets $(\alpha^\prime,\beta^\prime)$, $(\alpha,\beta)$ define the exponential decay characteristic of single body interphase, such as the extent of interphase and maximum value. Parameters $(\eta^\prime,\zeta^\prime)$ and $(\eta,\zeta)$ are, respectively, scaling and offset values for compound effect. Figure \ref{singlebody_decay} and Figure \ref{compounding} shows interphase images of single body effect and multi-body compound effect reconstructed by above computations for normalized property profile for two RVE samples from the DoE. The parameter set $(\alpha, \beta, \eta, \zeta)$ was inferred from the AFM modulus profile from a typical experiment and $(\alpha^\prime, \beta^\prime, \eta^\prime, \zeta^\prime)$ are inferred using fluorescence microscopy data from literature \cite{min2019}

\subsection{Elastoplastic Interphase material Model}\label{El_Pl_Interphase}
\begin{figure}
    \centering
    \includegraphics[width=\textwidth]{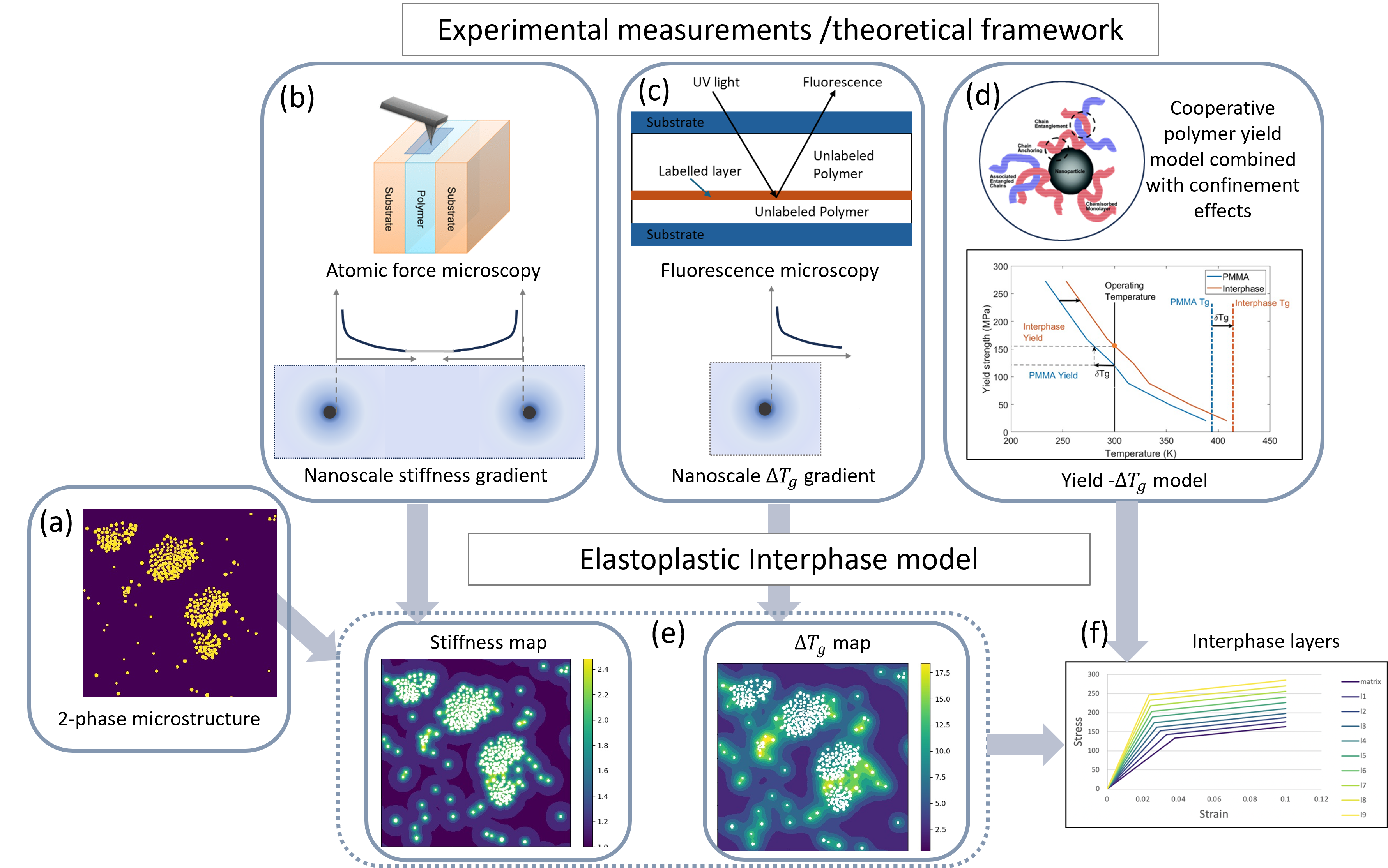}    
    \caption{ A flowchart demonstrating how interfacial gradients for elastoplastic property are constructed by combining stiffness and yield gradients. (a) binary-phase micro SVE with matrix(purple) and particle dispersion as well as agglomerates(yellow) (b) fluorescence spectroscopy measurement data \cite{min2019} used to derive gradient decay parameters for (c) local interfacial gradients that are transformed into local yield gradients using (d) temperature shift rule applied to macro-scale PMMA temperature-stress-strain data \cite{richeton2006influence}. (e) Direct local stiffness measurement from AFM nanoindentation experiments used to determine (f) local interfacial gradients for Young's modulus. (g) Local stiffness and yield gradients are combined using binning technique to construct bi-linear stress-strain curves used to describe the elastoplastic property for the matrix and finite numbe of interphase layers.} 
    \label{shift model}
\end{figure}

In the absence of local property data beyond the linear material regime for polymer nanocomposites, we combine data for local elastic modulus (refer Figure \ref{shift model} (e))  and local $T_g$ (refer Figure \ref{shift model} (b)) measurements on model nanocomposites as well as polymer's (here PMMA) macro scale elastic-plastic behavior to infer the effect of interfacial interactions on molecular mobility and glass transition temperature. Then we proceed to propose a material model for inelastic (stiffness and yield) properties of the interphase as demonstrated, step by step, in Figure \ref{shift model}. 

There have been multiple experimental studies that have tried to quantify the effect of operating temperatures on the macroscopic mechanical behavior of neat polymers \cite{bauwens1973compression,blumenthal2002influence,chen2002tension,cady2003determining,richeton2006influence}. These studies all point to the fact that the stress-strain behavior of macroscopic polymers under mechanical loading greatly depends on operating temperatures relative to their glass transition temperatures. It has been specifically observed for amorphous polymers that yield stress decreases and initial hardening in the nonlinear regime begins to diminish as operating temperatures approach the glass transition temperature. Thermally activated and cooperative models of yield process rooted in molecular theories have been proposed and refined through the cumulative work of multiple researchers over the years \cite{eyring1936viscosity, robertson1966theory,fotheringham1978role,richeton2005formulation}. Cooperative model states that the yield process involves the cooperative motion of chain segments governed by glass transition temperatures, $T_g$, through molecular mobility. At temperatures below $T_g$, the polymer chains do not have enough thermal energy and mobility to undergo cooperative movements and act stiffer, recording higher yield stress values but exhibiting brittle failure. On the other hand, at temperatures approaching or above $T_g$, thermal energy, and mobility can afford cooperative chain movements and record lower yield values along with more ductile failure and high toughness.

We use this understanding to propose a temperature shift model (refer Figure \ref{shift model} (c)) for local yield stress that can be used to derive interphase material properties based on polymer bulk property data. As explained by the cooperative model, the yield strength drops as the use temperature approaches the polymer glass transition temperature. If we assume attractive interfacial interactions, the polymer in the interphase region exhibits reduced chain mobility and overall reduction and delay in cooperative chain movements. This change would correspond to an increase in the extent of the linear regime and push the yield stress value upwards. Our key assumption is that the amount by which polymer yield stress can increase in the interphase region is related to the local change in $T_g$ of the interfacial region polymer, as indicated by an arrow of size $\delta T_g$ in Figure \ref{shift model} (c). Thus, the yield-temperature curve for interphase (orange curve in Figure \ref{shift model} (c)) can be obtained by shifting polymer curve (blue curve in Figure \ref{shift model} (c)) the same amount and direction corresponding to a change in $T_g$. Full field $\Delta T_g$ and Young's modulus are calculated first directly by gradient interphase equations \ref{singleE} through \ref{compound_delTg}, then full field yield strength can be determined by substituting full field $\Delta T_g$ values into the yield-temperature curve of polymer matrix and interphase. 

This interphase shift method was inspired by an approach used in viscoelastic and dielectric interphase modelling based on time temperature superposition principle (TTSP) \cite{brinson2008polymer,wang2018identifying,prabhune2023design}. Such temperature shift models have been shown to well represent interfacial confinement effects on local properties. Validity of TTSP in finite strain deformation regime have been established by multiple experimental and molecular dynamics simulation studies \cite{diani2015direct, xiao2013modeling,federico2018large,zhao2024time}.

Next, a bi-linear stress-strain model is utilized to define elasto-plastic properties in the interphase and polymer matrix area. Full field continuous property values are digitized into finite buckets of property levels called 'bins'. These bins approximate continuous gradients represented by equations \ref{singleE} through \ref{compound_delTg} through finite number of steps (refer to SI Figure \ref{SI_binning}). Young's modulus and yield stress from overlapping bins in stiffness and $\Delta T_g$ gradient maps are combined to form a bi-linear curve, as depicted in Figure \ref{El_Pl_Interphase}. 

\subsection{Characterization and reconstruction of the microstructure with agglomerated particles}\label{mcr}
Microstructure characterization and reconstruction plays a crucial role in understanding the system \cite{bostanabad2018computational}. For the system of interest, the key microstructure feature is the presence of nanoparticle agglomerations. Therefore, the characterization and reconstruction revolve around describing the agglomeration features. The system can be characterized by the physical descriptors of total volume fraction (VF), number of agglomerations, and volume fraction of agglomerations present in the system. The difference in the total and agglomeration volume fraction describes the amount of isolated particles present in the system. Considering the characterization requirements, we implemented a sequential random adsorption inspired approach that involves adding particles to a new microstructure incrementally to reconstruct the material system. We considered a system size of 1 $\mu$m × 1 $\mu$m and reconstructed it on a 400 × 400 pixel image. Given total VF and agglomeration VF, the reconstruction process starts by randomly picking a location within the image as the center of agglomeration. Next, particles, sampled from a N(8,2) pixel distribution, are added sequentially around the center location within a pre-specified distance to form the agglomerations. An evolutionary algorithm is implemented to optimize the agglomeration sizes for the desired volume fractions.  To obtain realistic agglomerations the center of the agglomeration is tweaked occasionally to reconstruct non-circular shapes. Finally, for the remaining VF, difference between total VF and agglomeration VF, isolated particles are added randomly into the image. Example of reconstructions with specified characteristics are given in Fig.2. Additionally, to gain deeper insight into the underlying relationship between the microsturcture, interphase schemes of PNCs, and their elastoplastic response, we define and calculate three key descriptors for a PNC system: I-filler, weighted average of interphase property ($W_{int}$), and mean nearest neighbor distance (mNND). A brief explanation of these descriptors are provided below in Table \ref{descriptors_intermediate}.\\

\begin{table}[h!]
\label{descriptors_intermediate}
\centering
\caption{Definition and description of descriptor studied}

\begin{tabular}{ p{2cm}|p{5.8cm} p{5.8cm}  }
 \hline
 Descriptor& Definition  &Implication \\
 \hline
 I-filler   & Total perimeter of filler phase    &Amount of particle surface that is exposed to matrix phase and leading to the formation of interphase.\\
 \hline
 Weighted amount of interphase, $W_{int}$&   Summation of E in a microstructure with interphase for each pixel subtracted by summation of E in no interphase scheme for each pixel for the same microstructure  & Relevant extent of property increased in the whole system due to single body effect and compound effect\\
 \hline
 Mean nearest neighbor distance &The mean of all particles/ agglomerates’ nearest neighbor distance base on edge-to -edge distance & Crowing level of the system.\\
 \hline
\end{tabular}
\end{table}

\subsection{Design of Experiments}\label{DoE}
A DoE is devised that spans across microstructure parameters as well as interphase parameters. Microstructural part of the DoE serves the purpose of varying system dispersion whereas interfacial schemes model three possible qualitative pairings between property gradients for $E$ and $\sigma_y$. The DOE is shown in Table \ref{table_doe}, where note that five replicates of each configuration, called statistical volume element (SVE) are simulated due to the statistical nature of the geometric reconstructions. The details of how each of the parameters are varied are described below. 

\subsubsection{Microstructure Space} \label{mcr_space}
\begin{figure}
    \centering
    {\includegraphics[width=1\linewidth]{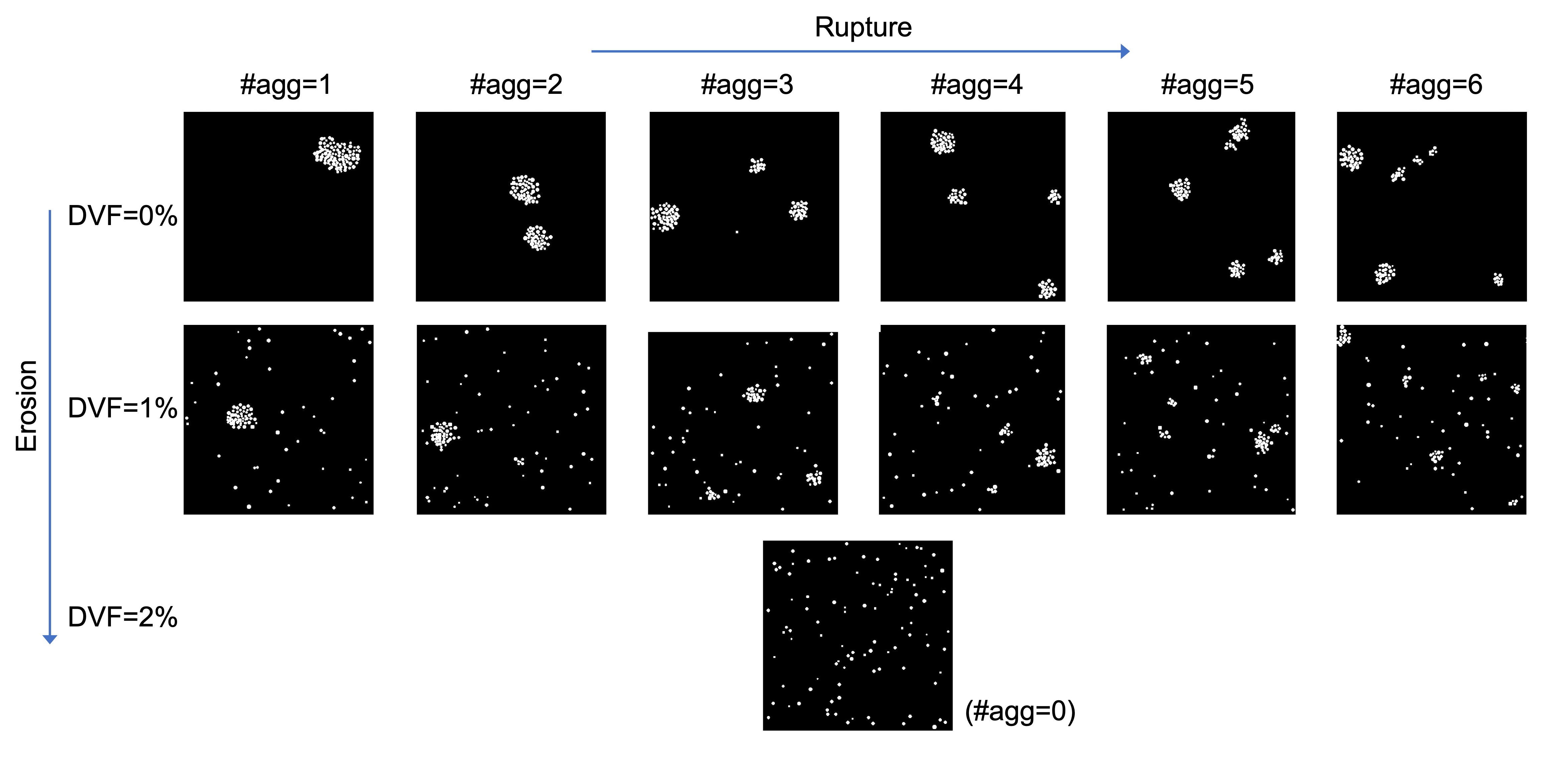}}
    \caption{reconstructed microstructures for total volume fraction 2\% demonstrating the concepts of erosion of agglomerates into individual isolated particles and rupture of large agglomerates into smaller, distributed ones, forming two axes of our microstructural variability. DVF refers to dispersed volume fraction and \#agg refers to the number of agglomerates in each image.}
    \label{microstructure_DoE}
\end{figure}

In this study, we created a design of experiment (DOE) space of varying microstructure characteristics. Specifically, we considered three total volume fraction (TVF) cases of 2\%, 5\%, and 8\%. Within each TVF level, we incrementally varied the dispersed volume fraction (DVF) from zero to respective TVF and adjusted number of agglomerations (nAggl) from 1 to 6. With a constant TVF and DVF, increasing number of agglomerations represent the \textbf{rupture} process, where a bigger agglomerate is broken into two or more smaller agglomerates. On the other hand, with a constant TVF and nAggl, increase DVF represents \textbf{erosion} where an agglomerate is eroded to create more disperse isolated particles rather than aggregates like agglomerations.
An example of reconstructed cases for the 2\% TVF are presented in Figure \ref{microstructure_DoE} with erosion and rupture directions. A similar approach was implemented for 5\% and 8\% TVF, with 4-5 replicate images for each configuration, leading to a total of 461 reconstructed images. The binary microstructure property setup remained consistent across cases: the Young's modulus $E$ of the matrix and filler are 3.5 GPa and 70 GPa, respectively, while the yield stress for the matrix $(\sigma_y^{poly})$ is 132.9 MPa, and the filler is presumed to behave linearly in the small-strain region. 

\begin{table}[htb]
    \centering
    \caption{Design of experiments with specific values for all design variables. All possible combinations of below design variable values are simulated. For instance, for a total VF 5\%,  DVF is varied from 0\% to 5\% with simultaneous variation in agglomerated VF going from 5\% to 0\%, at each combination of VF and DVF, number of agglomerations are varied from 0 to 6. For total VF 8\% , DVF is varied from 0\% to 8\%}
 \begin{tabular}{ll}
\toprule
 Design Variables & Range \\ 
 \midrule
Total volume fraction, VF & [2,5,8]\% \\
Number of agglomerations, nAggl  & [0,1,2,3,4,5,6] \\ 
Volume fraction of isolated particles, DVF & [0,1,2,3,4,5,6,7,8]\% \\ 
Volume fraction of agglomerated particles, AgglVF & [0,1,2,3,4,5,6,7,8]\% \\  
\bottomrule
    \end{tabular}
    \label{table_doe}
\end{table}

\subsubsection{Interphase space}
Four different interphase conditions are considered in our DOE. From the "no interphase" case to the most accurate case of a full gradient interphase. The details of the assumptions for these configurations are described in detail below. 
\begin{enumerate}
    \item \textbf{Coupled gradient interphase} Here we assume that an identical interfacial molecular mechanisms produces both local property changes in elastic stiffness and local glass transition. Hence, we use a same set of parameters for describing both stiffness gradient and local glass transition temperature variation. To account for different polymer-particle systems, we consider two types of gradient decay, slow and fast, as shown in the schematic in Figure \ref{decay}.
    \begin{figure}
        \centering
        \includegraphics[width=0.75\linewidth]{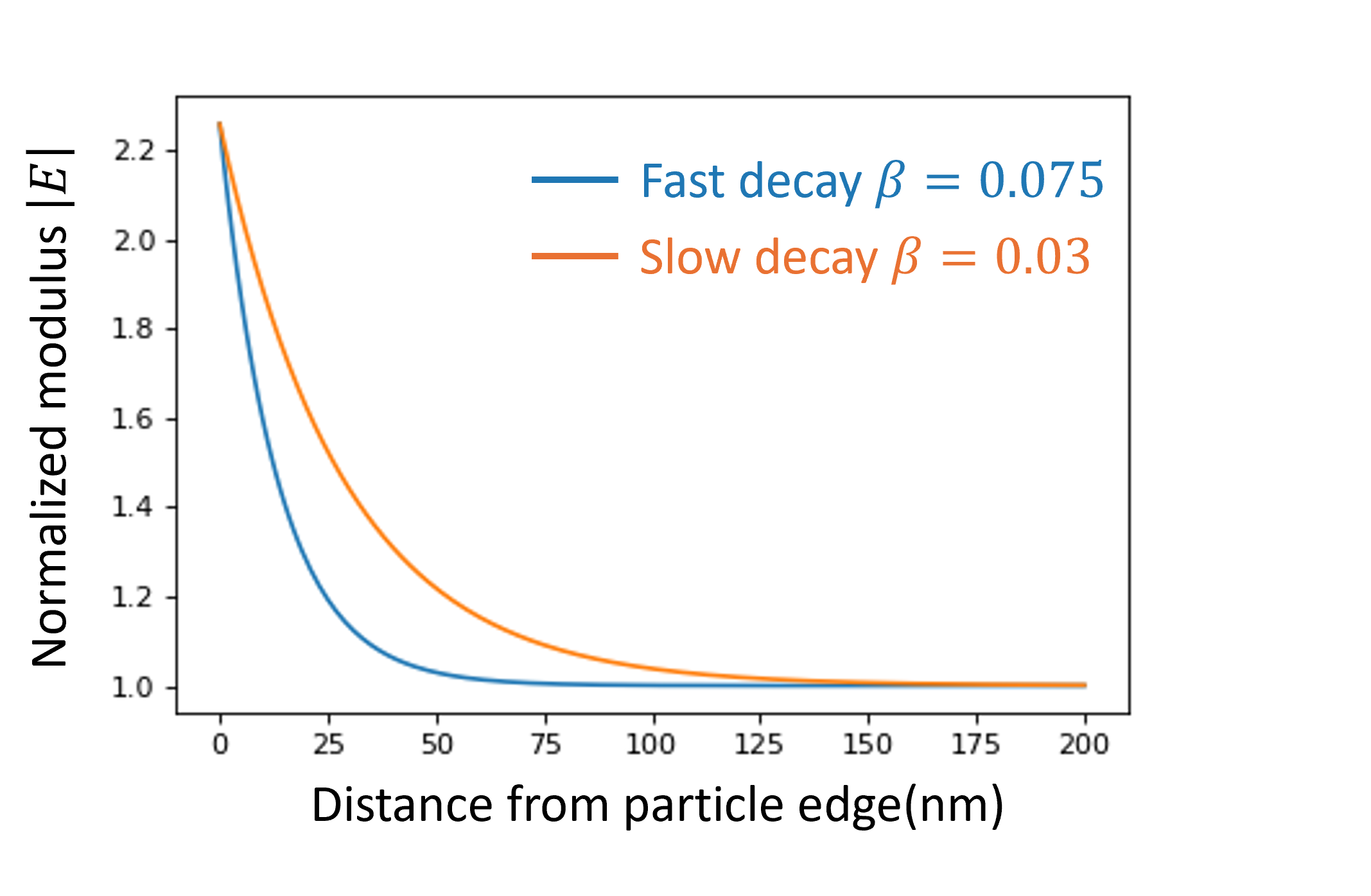}
        \caption{Illustration of two types of gradient interphases, representative of two different polymer-particle systems. A slow decay represents a system with strong polymer - particle interactions resulting in an extended interphase regime. A fast decay represents a system with weaker polymer-particle interactions and correspondingly diminished magnitude and extent of interphase regime}
        \label{decay}
    \end{figure}
    Slow decay rate of $\beta = \beta' = 0.03$ was inferred from fluorescence spectroscopy measurements data from \cite{zhang_determination_2018} for $T_g$ gradients on model nanocomposites. Fast decay rate of $\beta = \beta' = 0.075$ was inferred from elastic modulus profile measured by AFM on model composites from \cite{zhang_stiffness_2017,LI2019100277}. The modulus and yield stress gradient maps are computed as described in Section \ref{El_Pl_Interphase} and shown in Figure \ref{shift model}. The continuous property values are digitized into 10 bins (including polymer matrix) for FEA implementation. Top two rows in Figure \ref{E_Erosion_gradients} and \ref{sigma_Erosion_gradients}, in that order, display of modulus and delta Tg naocomposite maps for the slow and fast decay coupled interphase in case of erosion. Coupled gradient interphase have the advantage of shared bins for $\Delta T_g$ and $E$, which reduces the number of input layers required for FEA.\\
    \item \textbf{Decoupled gradient interphase} Experimental studies on effects of $T_g$ and modulus confinements in thin films have suggested length scale differences in $T_g$ and stiffness gradients \cite{torres2009elastic,ye2015understanding,askar2016stiffness,zhang_stiffness_2017}. To achieve more precise modeling, we developed a decoupled gradient interphase scheme, allowing modulus and delta Tg to have different decay rates based on experimental observations \cite{zhang_stiffness_2017,min2019}. Specifically, the modulus follows a faster decay rate, same as the value in the fast coupled gradient, while $\Delta T_g$ adheres to a slower decay rate seen in the slow coupled gradient. As a result, the local modulus and yield stress are independent and form distinct bins, as Figure \ref{shift model} shows. In this configuration, more than 50 combination of digitized stiffness and Tg are formed, increasing the complexity of the model.\\
    \item \textbf{uniform interphase} While we know from experimental data that the polymer properties decay in a gradient fashion, away from each filler surface, conducting FEA simulations with gradient interphase is computationally intensive and resource-demanding, especially in 3D. Therefore, we also consider a simplified uniform interphase representation. Properties for uniform interphase are calculated individually for each SVE, by equating weighted property value for gradient map to uniform map. The width of uniform interphase is fixed at 20 pixels (50 nm). \\
    \item \textbf{Without interphase} We include also the simplest two phase system, just matrix and particles, without any interphase as a benchmark case. This case helps to qualitatively decouple effect of interphase from the effect of particles and their pure morphology.
\end{enumerate}
\begin{figure}
    \centering
    \includegraphics[width=\textwidth]{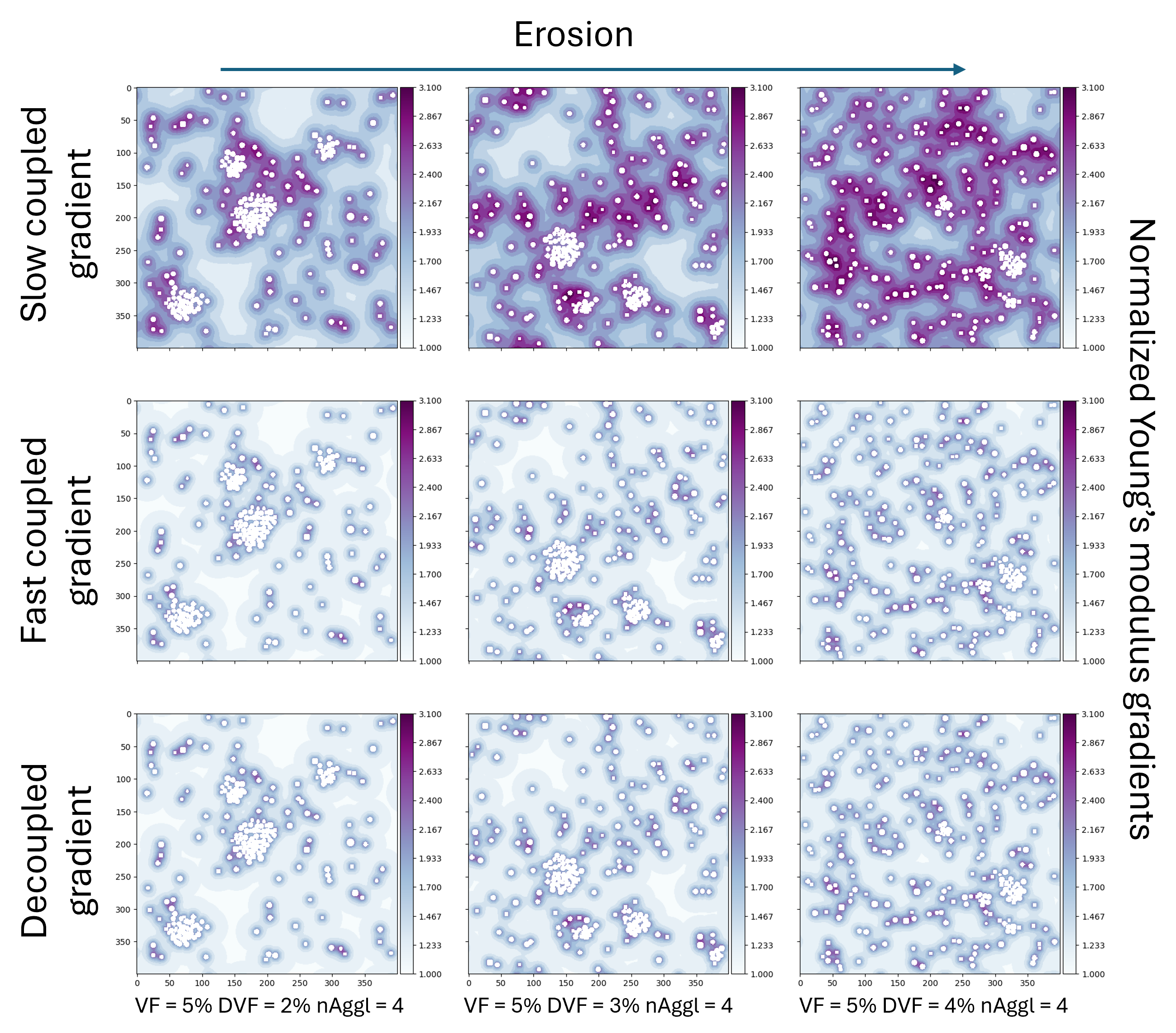}
    \caption{Young's modulus interfacial Gradients for three interphase schemes in the case of erosion}
    \label{E_Erosion_gradients}
\end{figure}
\begin{figure}
    \centering
    \includegraphics[width=\textwidth]{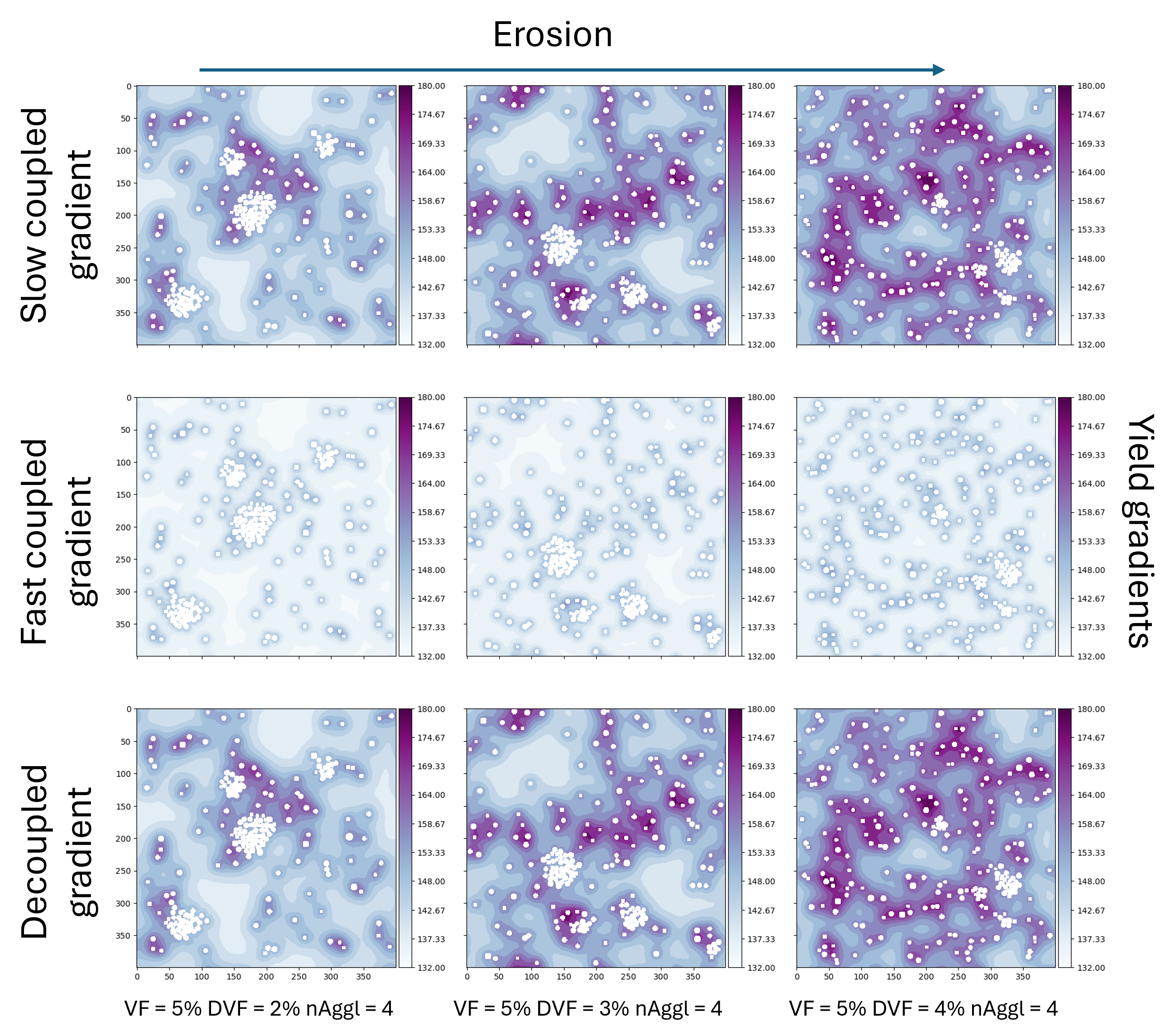}
    \caption{Yield stress interfacial Gradients for three interphase schemes in the case of Erosion}
    \label{sigma_Erosion_gradients}
\end{figure}

\subsection{Finite Element Simulations}
Each data point in design space(around 461 microstructure SVEs with 4 interphase schemes) was simulated in 2D Abaqus standard iterative solver under qusistatic strain loading of 10\% in order to experience yielding and plastic deformation. Each SVE FE model consisted of 1600 2D plain strain elements. We employed kinematically periodic BCs using '*Equation' keyword in abaqus input file. Periodic BCs implied that the single micron size SVE simulates effective stress-strain behavior and hence material properties of a nanocomposite bulk that can be reconstructed by repetition in both the axes. Abaqus built-in elasto-plastic material model was used. Composite macro stress-strain curve was retrieved from abaqus output file by calculating domain average stress and strain at each load step point. Macros scale effective properties $E^{macro}$ and $\sigma_y^{macro}$ are calculated using 0.2\% offset method, which is a standard engineering practice. More details included in SI. 

\section{Results} \label{results}
The full DOE is simulated in FEA as described above to obtain stress-strain response curves of the each composite RVE. The nanocomposite modulus $E^{macro}$ and yield strength $\sigma_y^{macro}$ are extracted from the simulated stress-strain data. Property values are averaged across the replicates within each DoE data point. The slope of the linear region is computed as $E^{macro}$, while the 0.2\% permanent set method is used to identify $\sigma_y^{macro}$. Graphical illustration and details are included in the Supporting Information. The results for the simulation of the DOE are presented exploring the impact of erosion (distributing particles from agglomerations randomly throughout the bulk) and the impact of rupture (breaking large agglomerations into more numerous small agglomerations).

\subsection{Effect of erosion process}
To understand the role of the erosion process on elasto-plastic properties, $E^{macro}$ and $\sigma_y^{macro}$ are further averaged across number of agglomerations and plotted against DVF, individually for TVF values of 2\%, 5\%, and 8\%. Figures \ref{erosion_E} and \ref{erosion_yield} compare the $E^{macro}$ and $\sigma_{y}^{macro}$ performance of the different gradient interphase configurations within each TVF subgroup. In each plot, advancing along the x-axis indicates increasing erosion, while error bars represent the impact of rupture.

\subsubsection{Young's Modulus}
\begin{figure}
    \centering
    \subfigure[]{\includegraphics[width=.32\linewidth]{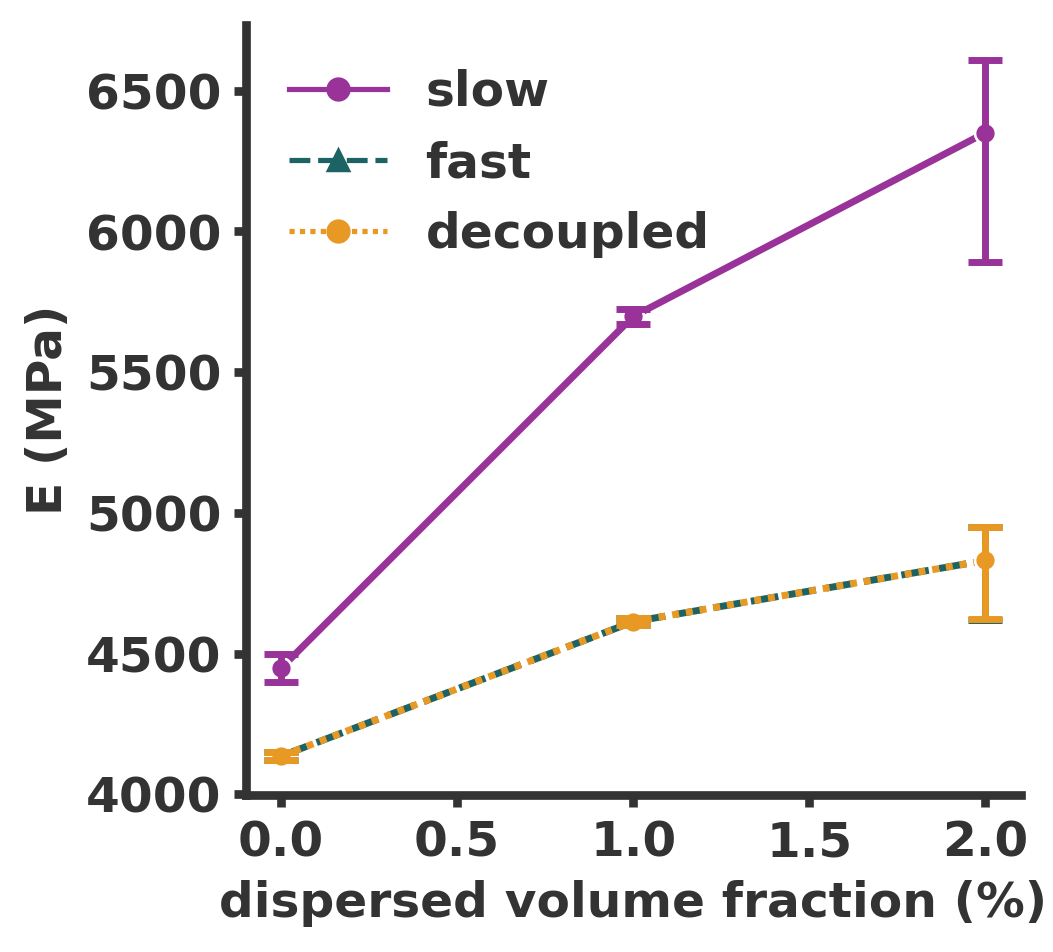}}
    \subfigure[]{\includegraphics[width=.32\linewidth]{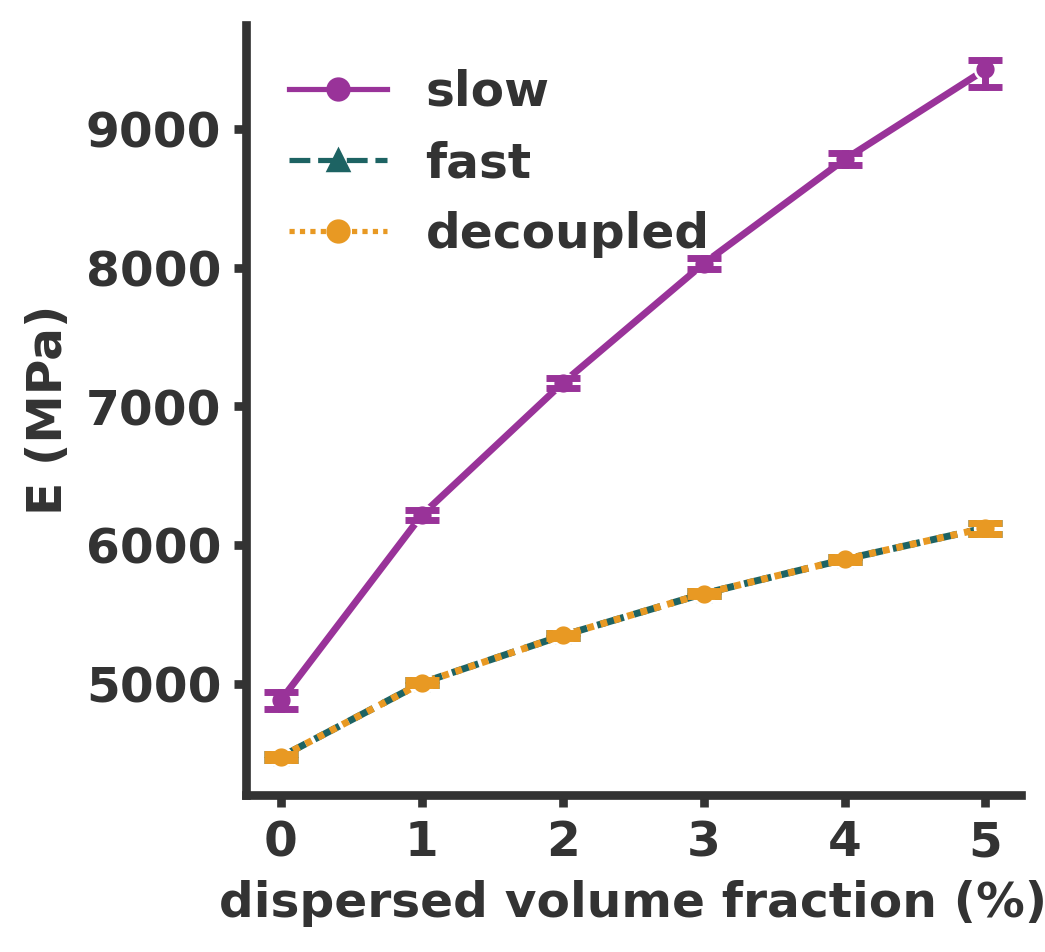}}
    \subfigure[]{\includegraphics[width=.32\linewidth]{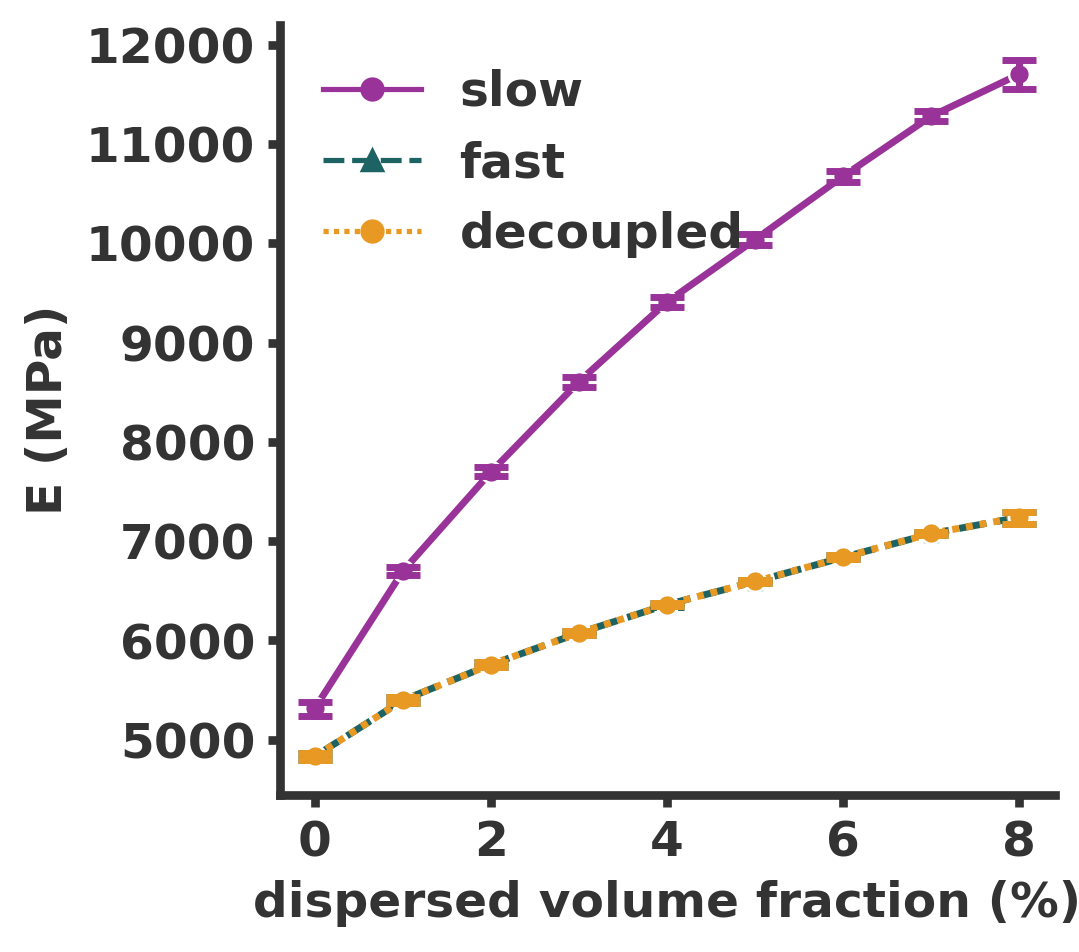}}
    \caption{Young's modulus with erosion: DVF (dispersed volume fraction) relation for total volume fraction (TVF) cases (a) 2\% (b) 5\% (c) 8\%. The different line colors indicate different types of interphase configuration. 
}
    \label{erosion_E}
\end{figure}
We observe a monotonic increase in effective Young's modulus of nanocomposites with increasing dispersed particle volume fraction as shown in Figure \ref{erosion_E}. This increase can be attributed to the higher interphase percolation as a result of the increased dispersion. Between interphase schemes, the slow gradient decay provides highest property values, owing to its largest extent of interphase percolation. Fast coupled and decoupled interphase schemes behave identically owing to the fact that they share gradient decay rate for $E^{micro}$. By comparing the levels of $E^{macro}$, among three levels of TVF, changes in DVF have a slower impact on modulus for the lower total volume fraction cases. For example, considering the slow decay case at 2\% and 8\% TVF, an increase from 0 to 2\% DVF increases modulus for the 8\% TVF case by \~50\% and just under 40\% for the 2\% TVF case (see red circles in Fig 7.). Thus, in order to achieve higher modulus, improved dispersion is more critical in systems with lower particle loading than those with higher, due to accelerated interphase connectivity with higher loading (Figure \ref{macroE_noIntph_Erosion}). The error bars due to the variation in the number of agglomerations and the agglomerated volume fraction are minimal, except for the cases of fully dispersed $2\%$ TVF systems. We attribute this variance to higher interphase percolation variability due to low particle loading .  
\subsubsection{Yield strength} \label{erosionYld}
\begin{figure}
    \centering
    \subfigure[]{\includegraphics[width=.32\linewidth]{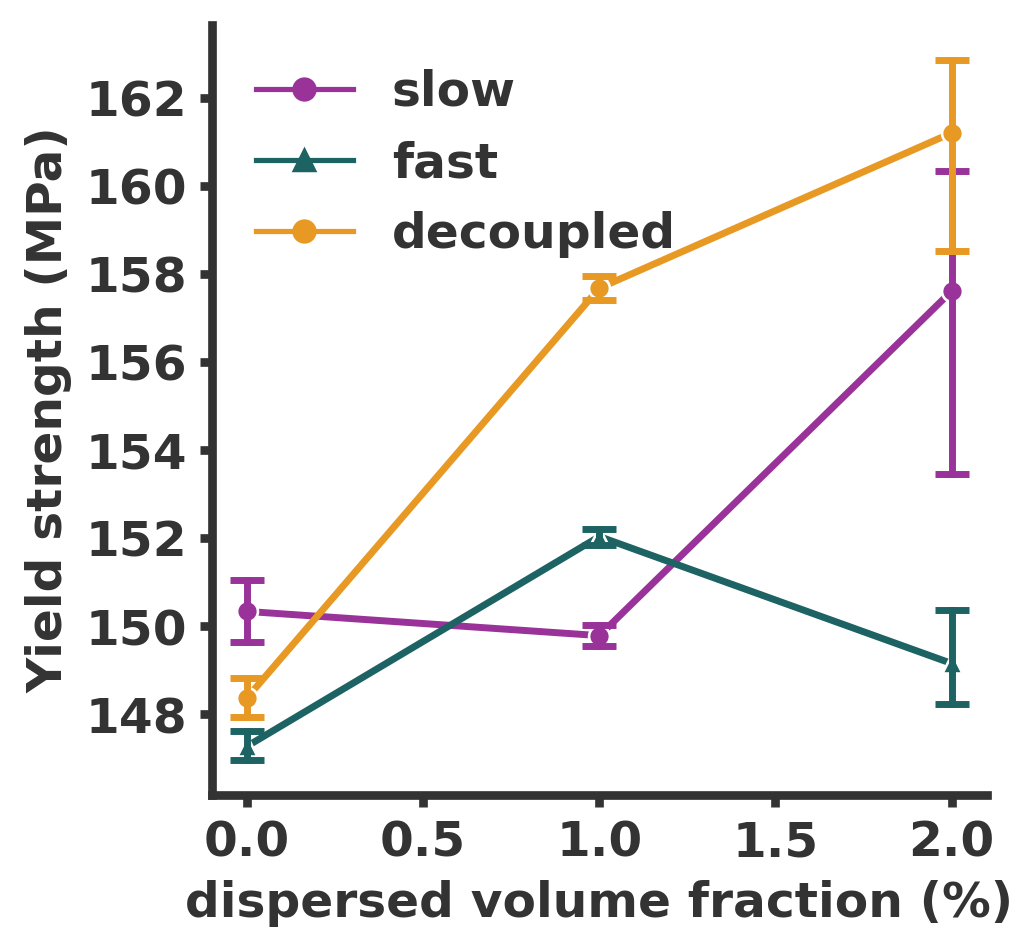}}
    \subfigure[]{\includegraphics[width=.32\linewidth]{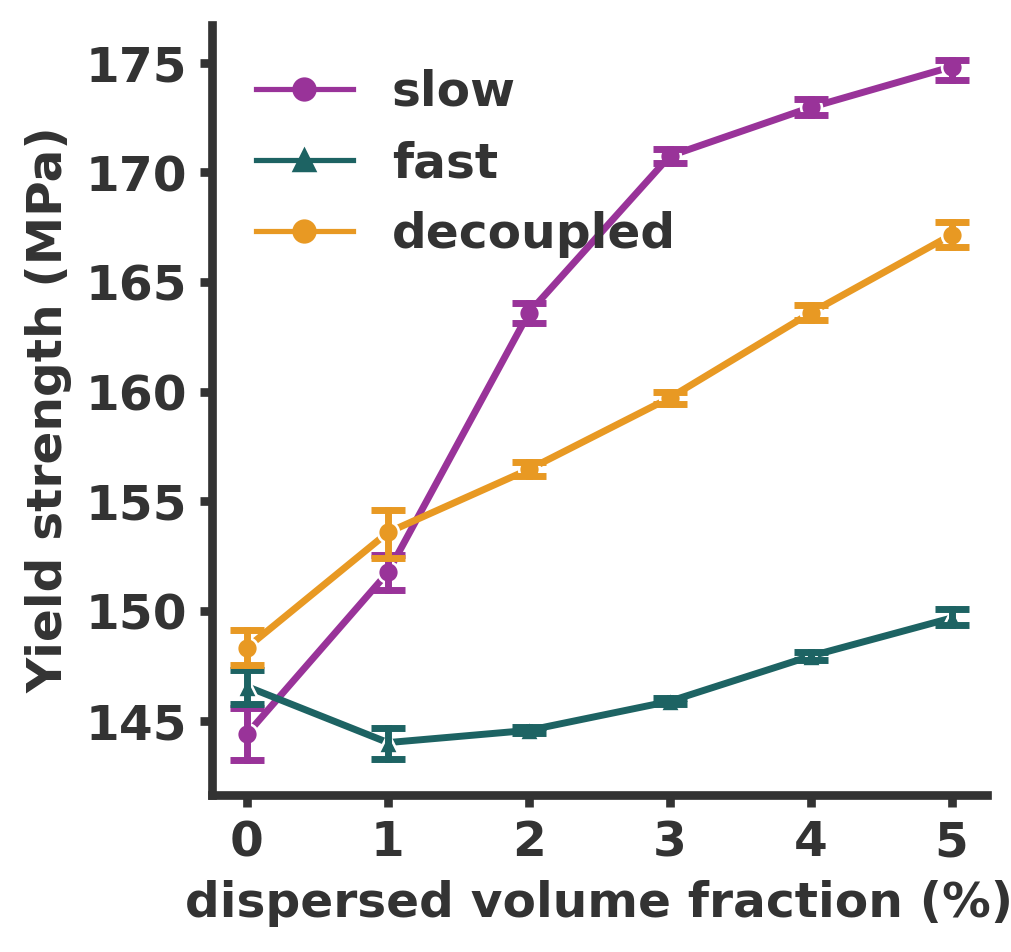}}
    \subfigure[]{\includegraphics[width=.32\linewidth]{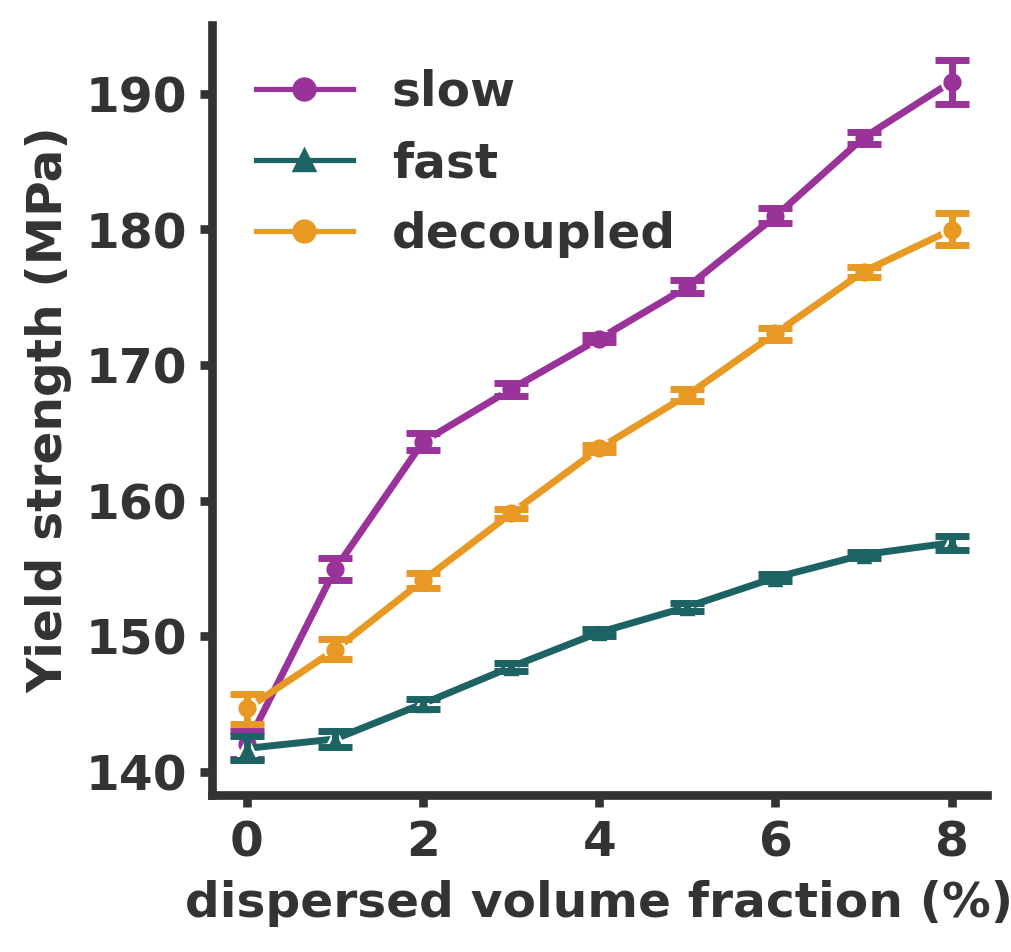}}
    \caption{Yield Strength with erosion: DVF (dispersed volume fraction) relation for total volume fraction (TVF) cases (a) 2\% (b) 5\% (c) 8\%. The different line colors indicated different types of interphase configuration. The no-interphase case is not included here. }
    \label{erosion_yield}
\end{figure}
Similar to $E^{macro}$, we observe a monotonic increase in $\sigma_y^{macro}$ values with increase in dispersed volume fraction as shown in Figure \ref{erosion_yield}. However, unlike $E^{macro}$,  yield strength values for decoupled and slow interphase cases differ despite sharing an identical decay rate for interfacial gradients of local yield stress.  Mechanical yielding is a nonlinear, path dependent property that is governed by how local elasto-plastic deformations evolve. This deformation evolution is dictated by local yield gradients as well as local $E$ gradients.   

Further, for slow interphase decay, we see an interesting slow down in dispersion driven property enhancement for the case of TVF $5\%$ at DVF $3\%$ and for the case of TVF $8\%$ at DVF $2\%$. For fast interphase decay, the property enhancement slowdown is more subtle and gradual in the case of TVF $8\%$ and not observable for TVF $5\%$. The phenomena is absent for the case of decoupled interphase scheme where yield strength increases nearly linearly with dispersed volume fraction for all TVF cases. While it is tempting to attribute these findings to interphase percolation saturation, the difference in trends suggests that there is another phenomenon at play that governs the yielding mechanism. We identify this phenomenon to be the effect of local stress concentrations (LSCs). We discuss in Section\ref{yielding_mechanism} how local stress concentrations and local yield stress gradients compete to drive local yielding paths and macro yielding process processes.

The mixed trends observed in the lowest TVF case of $2\%$ and lower dispersion levels in TVF $5\%$ are believed to be due to the absence of interfacial compounding at lower dispersion levels. Inter-particle distances at which fast single body decay compounds are higher than those at which slow single body decay compounds. For the case of 2\%, we believe that combination of high systemic variability in terms of inter-particle distances and interphase percolations and the absence of compounding is behind the observed mixed trends. **For the case of the decoupled interphase, we can expect yield gradients to have compounded at lower dispersion levels whereas stiffness gradients may not have compounded. Thus, the balance will lean to the side of $\sigma_{y}^{micro}$ always for the decoupled interphase scheme. Hence, the decoupled interphase scheme exhibits approximately uniform increasing trends for $\sigma_{y}^{macro}$, throughout. **

\subsection{Effect of rupture process}
\begin{figure}
    \centering
    \subfigure[]{\includegraphics[width=.32\linewidth]{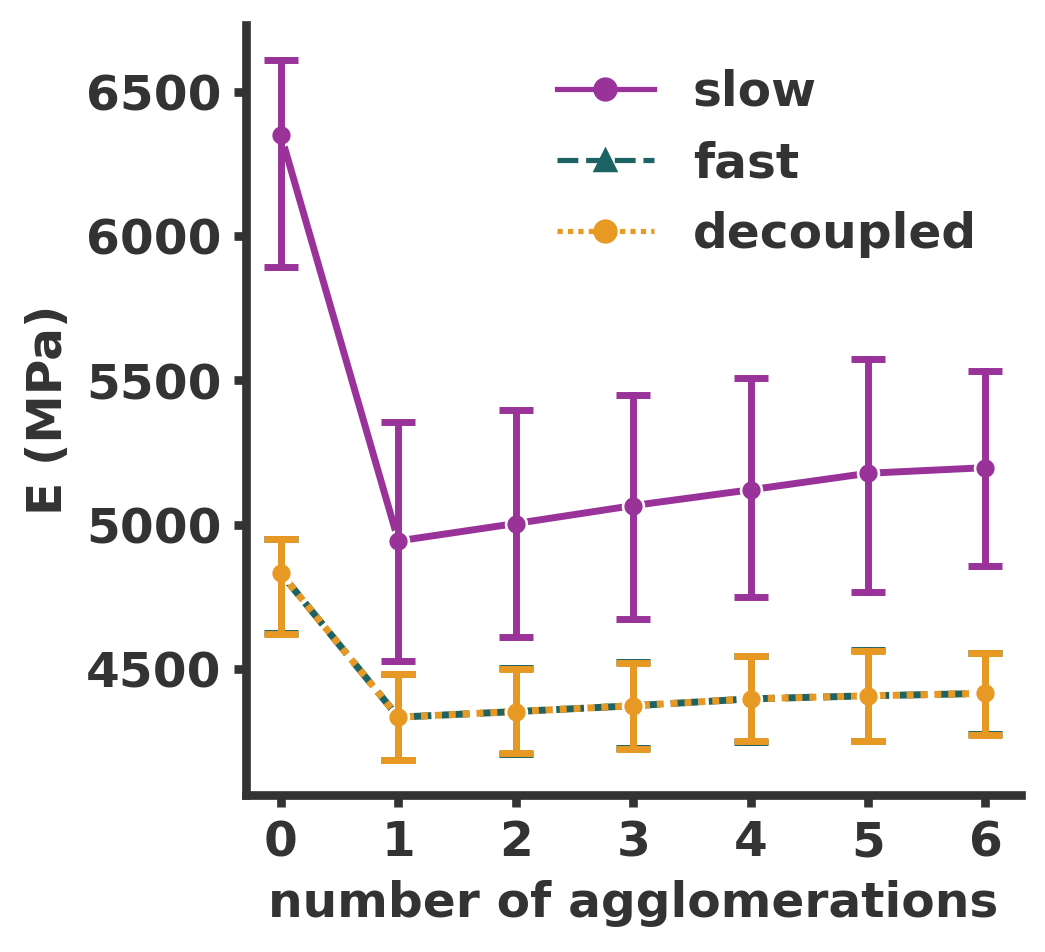}}
    \subfigure[]{\includegraphics[width=.32\linewidth]{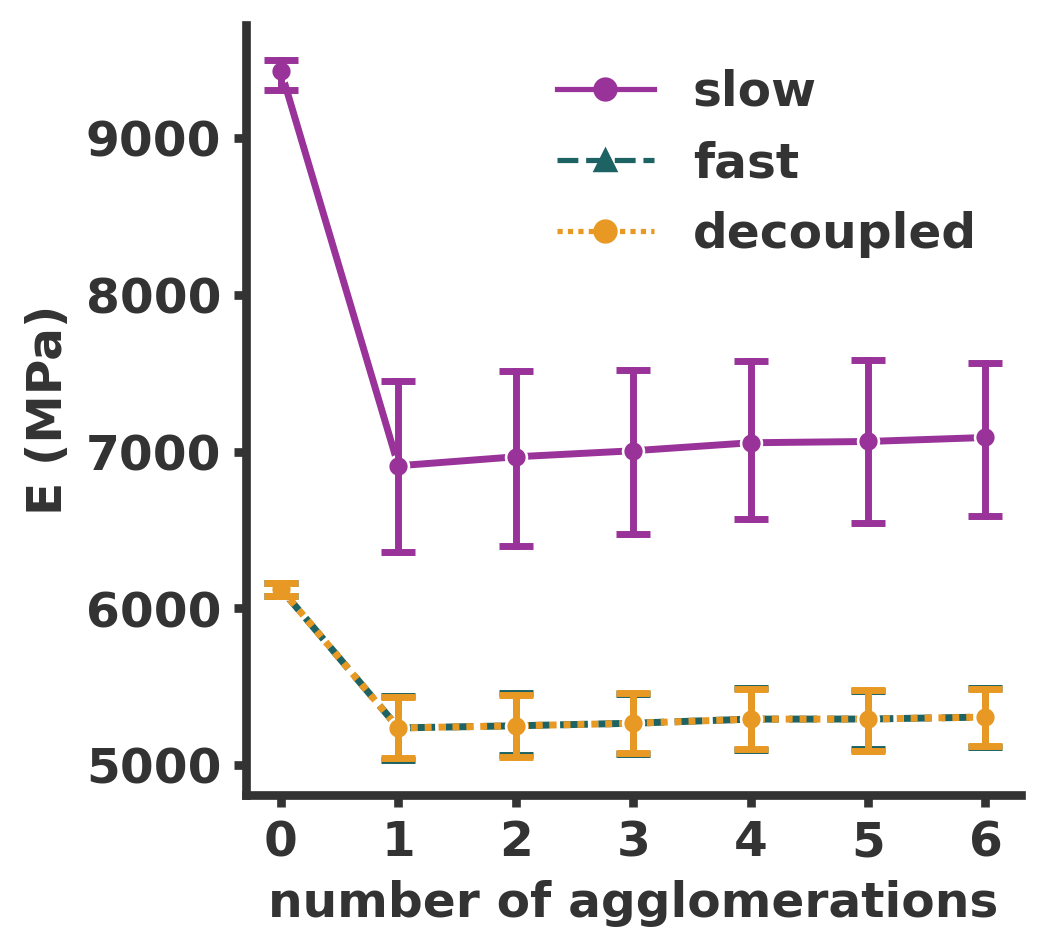}}
    \subfigure[]{\includegraphics[width=.32\linewidth]{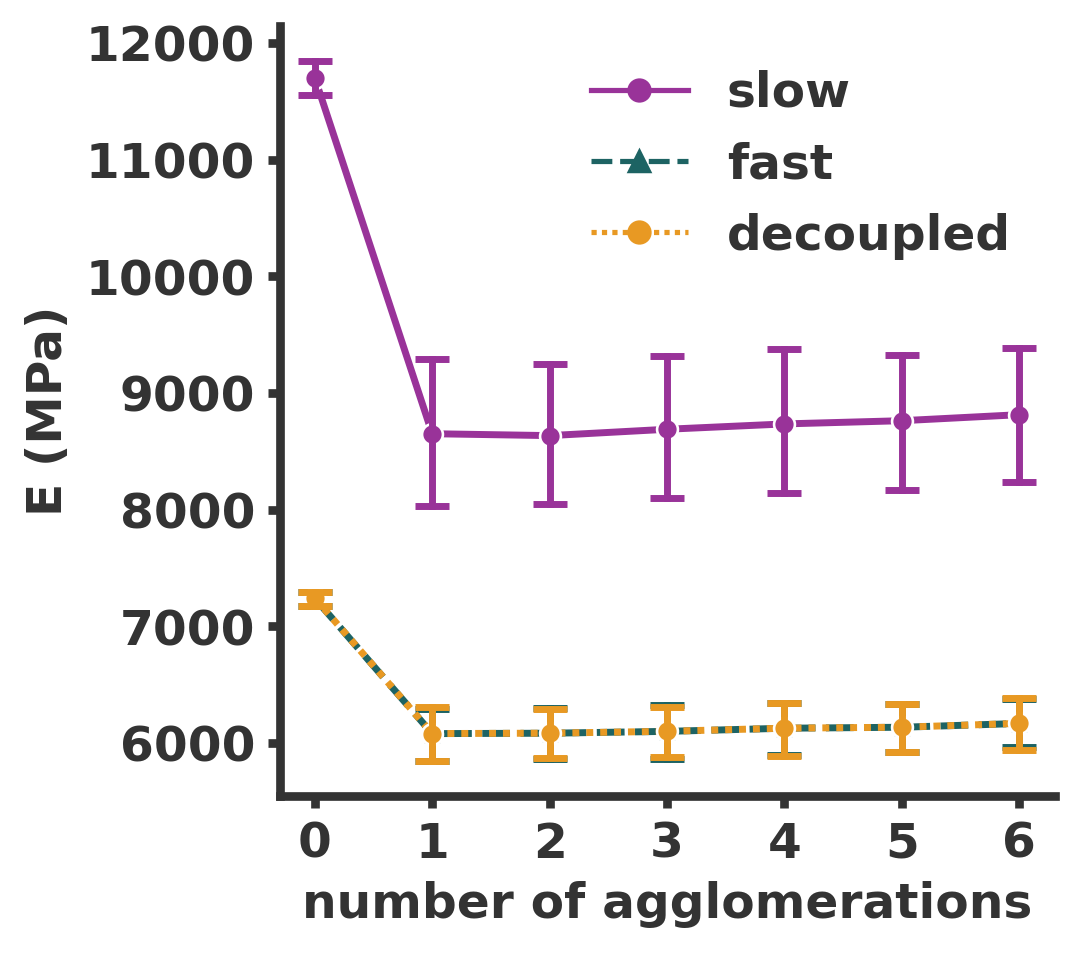}}
    \caption{Young's modulus with rupture: number of agglomerations relation for TVF (total volume fraction) cases (a) 2\% (b) 5\% (c) 8\%}
    \label{rupture_E}
\end{figure}

\begin{figure}
    \centering
    \subfigure[Total VF=2\%]{\includegraphics[width=.32\linewidth]{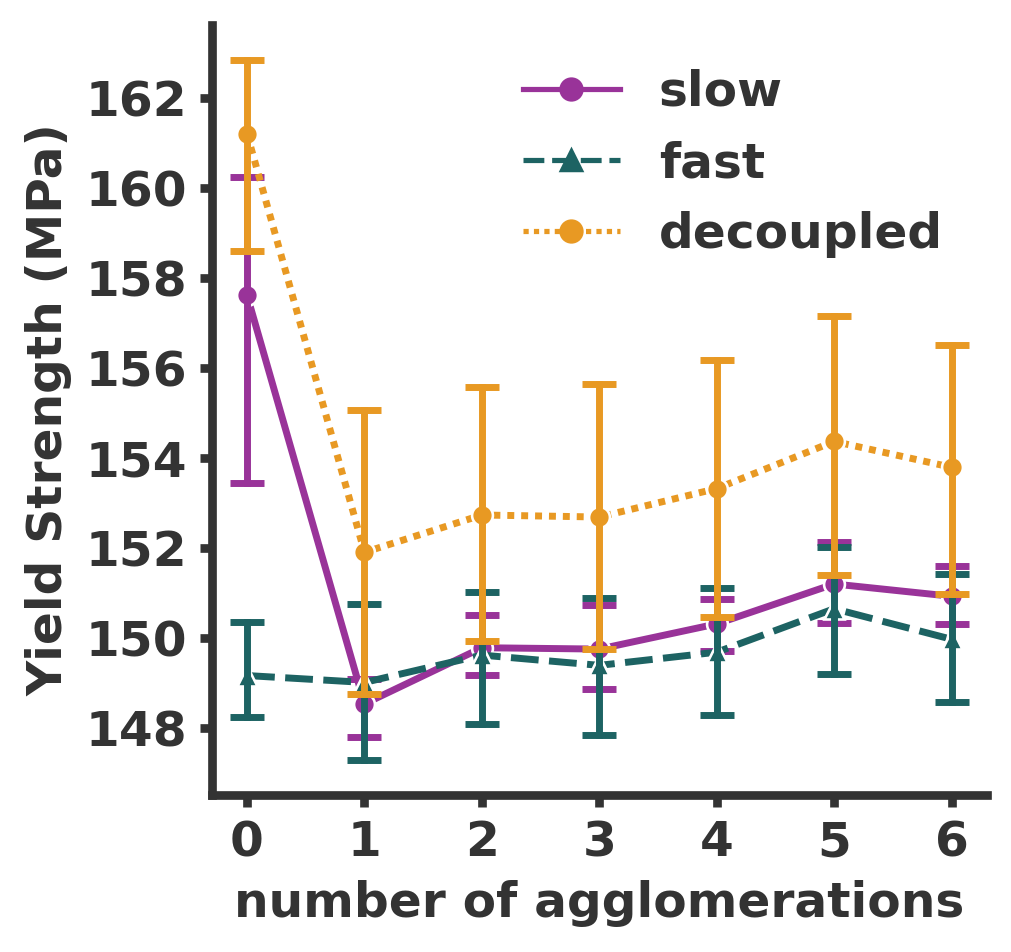}}
    \subfigure[Total VF=5\%]{\includegraphics[width=.32\linewidth]{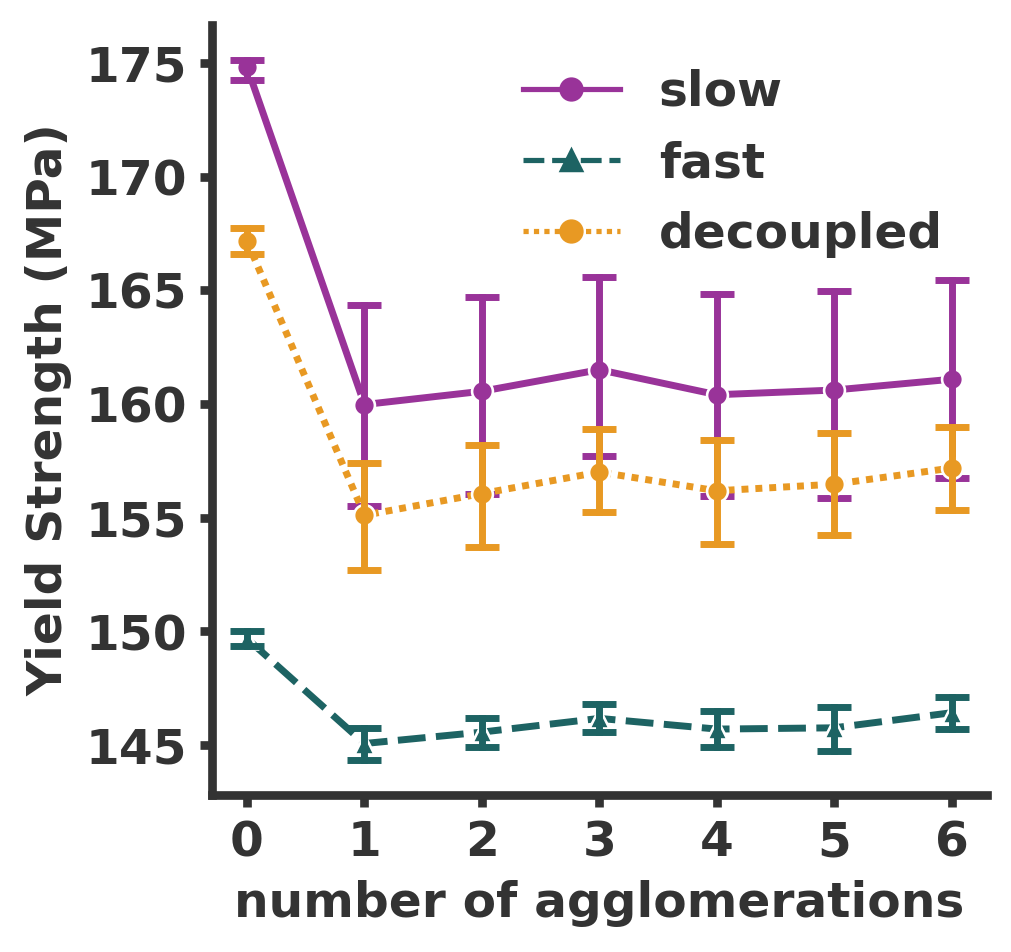}}
    \subfigure[Total VF=8\%]{\includegraphics[width=.32\linewidth]{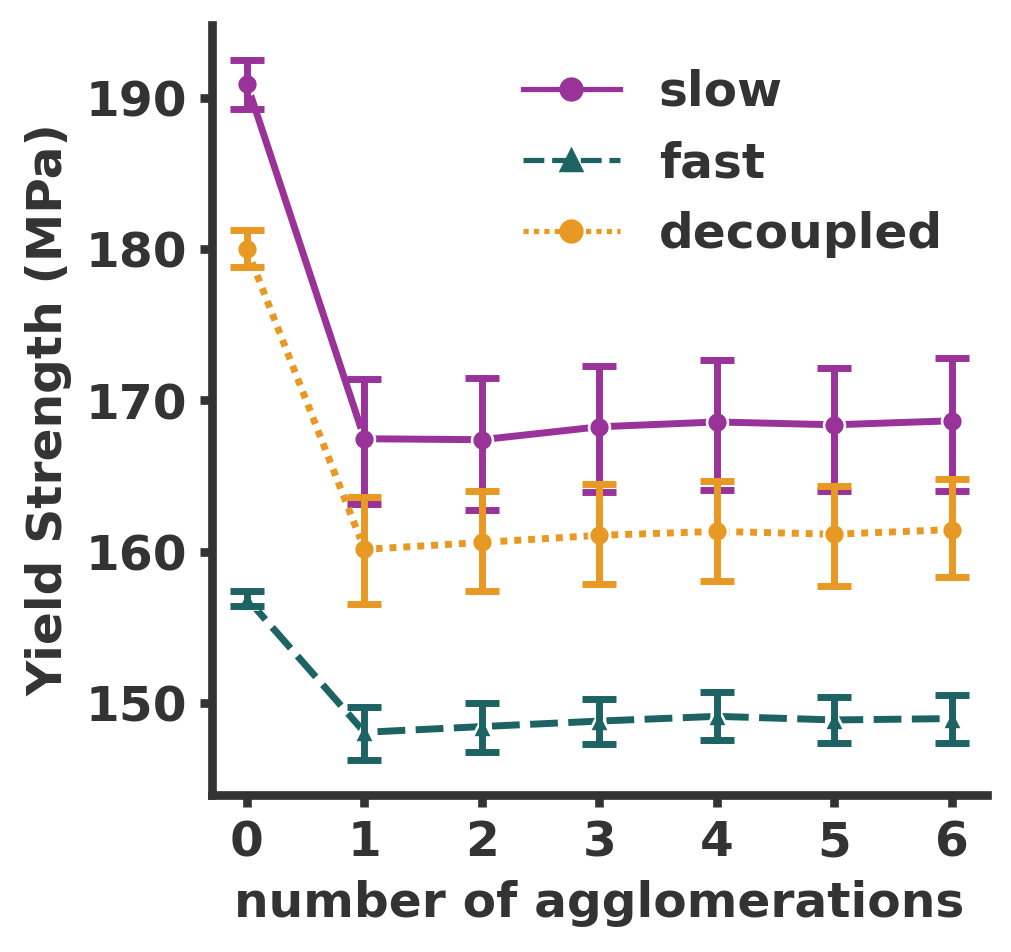}}
    \caption{Yield Strength with rupture: number of agglomerations relation for TVF (total volume fraction) cases (a) 2\% (b) 5\% (c) 8\%}
    \label{rupture_yield}
\end{figure}

To understand the role of the rupture process on elasto-plastic properties, $E^{macro}$ and $\sigma_y^{macro}$ is further averaged across DVF and plotted against number of agglomerations, individually for TVF values of 2\%, 5\%, and 8\%. Figures \ref{rupture_E} and \ref{rupture_yield} compare the $E^{macro}$ and $\sigma_{y}^{macro}$ performance of fast, slow decay coupled and decoupled gradient interphase within each TVF subgroup. In each plot, advancing along the x-axis indicates increasing rupture, while error bars represent the impact of erosion.

\subsubsection{Young's Modulus}
For the case of rupture, we observed that the enhancement in $E^{macro}$ values achieved by breaking larger agglomerations into smaller ones is limited compared to that brought on by the erosion process. This behavior can be attributed to the limited increase in dispersion, specifically interparticle distances, provided by the rupture process which prevents it from tapping into the effects of percolation and compounding to the same extent as erosion. We see, as expected, a complete overlap in decoupled and fast interphase schemes. Zero agglomerations is a significant maxiumum value and is shown for comparison to a fully dispersed case. In many material systems and fabrication methods, it is impossible to avoid agglomerates completely. In such cases, it is important to emphasize that higher number of smaller agglomerates is better for macroscale nanocomposite stiffness. Among attractive interfacial interaction induced interphase schemes, the extended interphase afforded by slow gradients are beneficial.  

\subsubsection{Yield strength} \label{ruptureYld}
Relative trends observed in $\sigma_y^{macro}$ in the rupture process are similar to that of the erosion process but with less significant increase in overall values. It is also noted that error bars across individual data point(average over DVF and replicates) are higher and are due to property increases due to increasing DVF.
\subsection{Microstructure descriptors as predictors of macroscale properties}
\begin{figure} 
    \label{prediction_of_E}
    \centering
    \subfigure[]{\includegraphics[width=.3\linewidth]{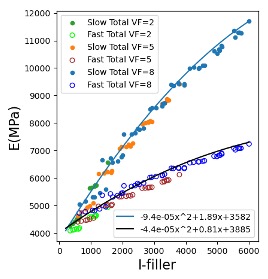}}
    \subfigure[]{\includegraphics[width=.3\linewidth]{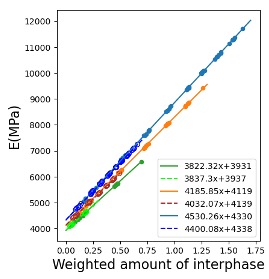}}
    \subfigure[]{\includegraphics[width=.3\linewidth]{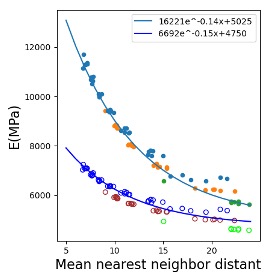}}
    \caption{Relation of I-filler(first column), weighted amount of interphase(second column), and mean nearest neighbor distance(Third column) with E(first row) and Yield stress(second row)}
\end{figure}

\subsection{Summary}
Increasing erosion becomes the primary factor in promoting stiffness for both fast and slow decay gradient interphase schemes. Increasing the DVF through erosion leads to significant enhancement in $E^{macro}$ and $\sigma_y^{macro}$. Meanwhile, rupture serves as a secondary factor, mainly influencing systems with few or no isolated particles. In such cases, highly ruptured systems exhibit slightly higher Young’s modulus, while systems with fewer and larger clusters display relatively lower modulus. Erosion and rupture both benefit from an interphase with a slow decay more than the fast decay due to its broader interphase range. The interphase area of eroding particles and clusters can sufficiently interact with each other increasing local stiffness as well as yield allowable through compounding.

To summarize, the DVF is the determining factor for stiffness, while agglomerated VF has a minor impact. The erosion of particles from agglomerates significantly promotes property improvement. Efforts to break agglomeration to smaller pieces are not particularly effective unless most of the particles are already clustered.

\section{Discussion}
In this section, we use the results of the simulated DOE to analyze the mechanisms of mechanical yielding in polymer nanocomposites with regards to interfacial gradients and particle dispersion levels. We identify two competing processes, namely local stress concentrations and interfacial yield gradients, that contribute to the yielding mechanism. Local stress distribution and local stress concentration is quantified by the full field von-mises stress, which is used in the yield criteria. In section \ref{dispersion_interphase_effect}, we explain the effects of dispersion, agglomerates and interfacial gradients on local stress concentrations. Then we analyze how these local field concentrations in conjunction with interfacial yield gradients shape the local yielding paths in section \ref{yielding_mechanism}. Finally, we build a yield susceptibility full field map as a predictor of yield percolation networks in Section 4.3.

\subsection{Effect of dispersion, agglomerates and interfacial gradients} 
\label{dispersion_interphase_effect}
To explain the observed levels of $E^{macro}$ and $\sigma_{y}^{macro}$ among three interphase schemes with respect to changing dispersion from the erosion and rupture processes, we consider the competition between local stress concentration zones (LSCZ) and interfacial yield gradients. The interfacial yield gradients are the gradients in the local yield stresses as described via the model in Section 2.2. 
\begin{figure}
    \centering
    \subfigure[without interphase]{\includegraphics[width=\linewidth]{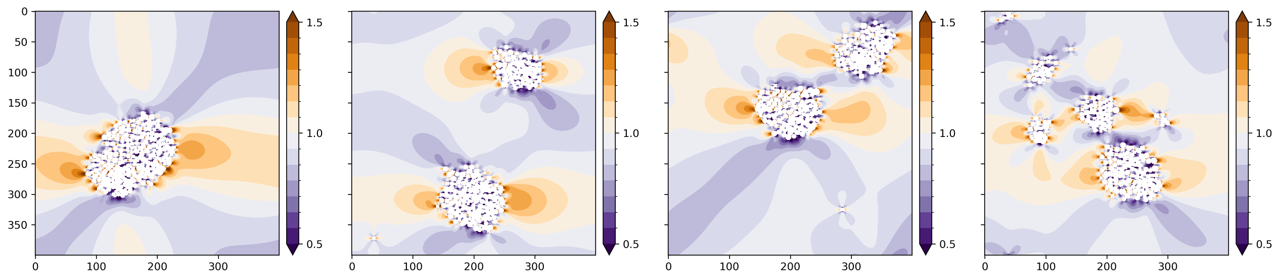}\label{LSCZ_rupture_no}}  
    \subfigure[with fast interfacial decay in $E^{micro}$]{\includegraphics[width=\linewidth]{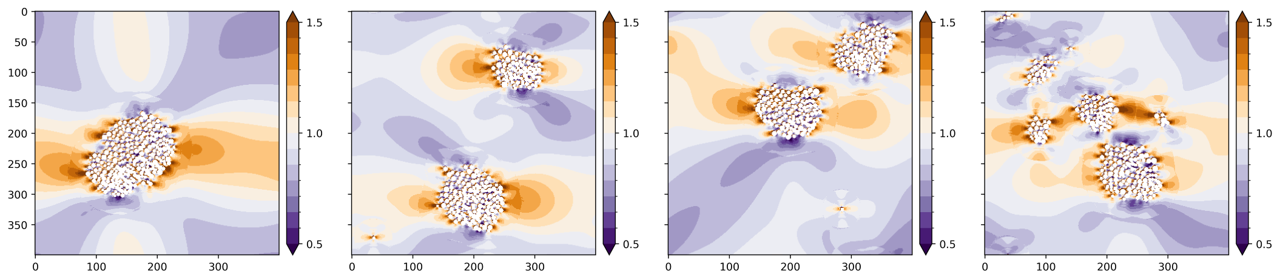}\label{LSCZ_rupture_fast}}  
    \subfigure[with slow interfacial decay in $E^{micro}$]{\includegraphics[width=\linewidth]{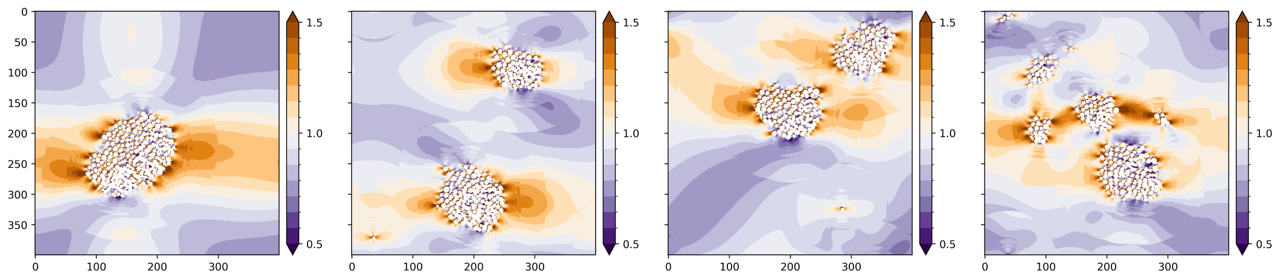}\label{LSCZ_rupture_slow}} 
    \caption{ Local stress concentrations (LCSZ) illustrated for the rupture process for the 5\% TVF case, with three different interphase schemes (rows a, b, c) and for differing numbers of agglomerations (columns going left to right). For the agglomerations we show a single agglomeration, then two instances of 2 agglomerations, and the 6 agglomerations. Local stress concentration factor(LCSF) is defined as usual with local stress relative to the macroscopic applied stress in the loading direction; LCSF magnitude is represented by the color bar, where white is equivalent to the applied macroscopic stress. 
 }
    \label{LSCZ_rupture}
\end{figure}

\begin{figure}
    \centering
    \subfigure[without interphase]{\includegraphics[width=\linewidth]{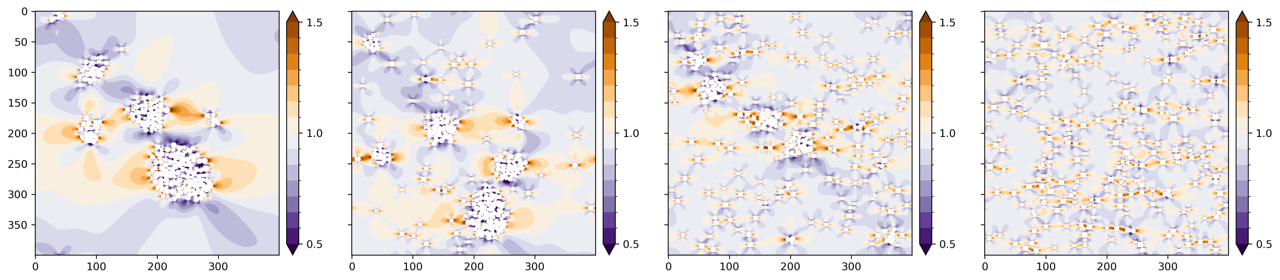}\label{LSCZ_erosion_no}} 
    \subfigure[with fast interfacial decay in $E^{micro}$]{\includegraphics[width=\linewidth]{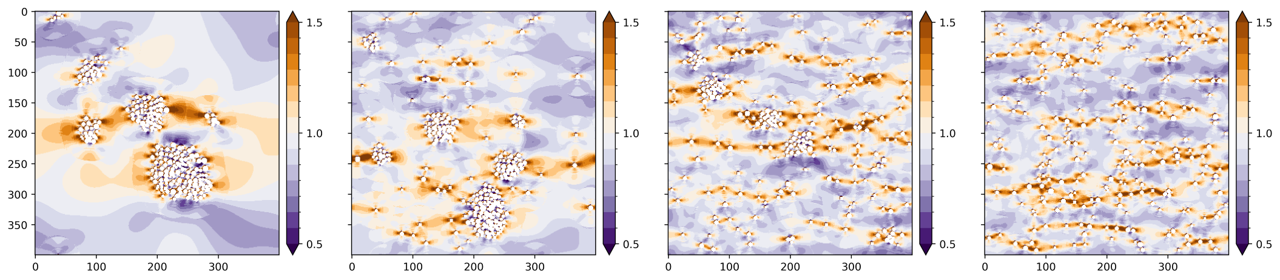}\label{LSCZ_erosion_fast}} 
    \subfigure[with slow interfacial decay in $E^{micro}$]{\includegraphics[width=\linewidth]{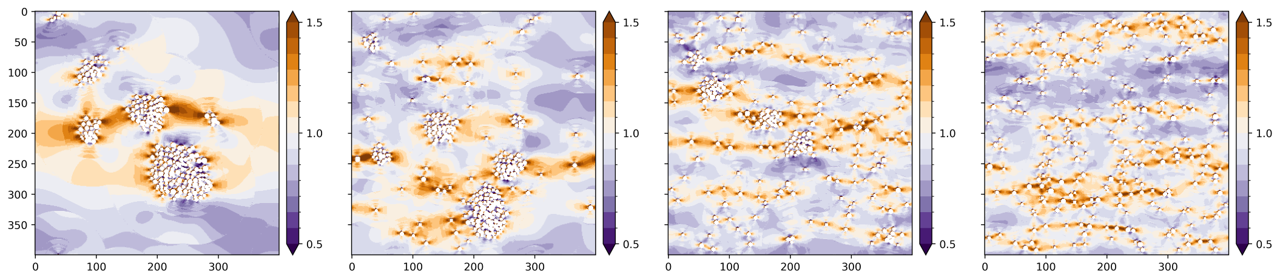}\label{LSCZ_erosion_slow}} 
    \caption{Local stress concentrations (LCSZ) illustrated for the erosion process for the 5\% TVF case, with three different interphase schemes (rows a, b, c) and for differing dispersed volume fraction (DVF) values (columns going left to right). For DVF we show the cases: 0\%, 1\%, 3\% and 5\% (no agglomeration). Local stress concentrations same as defined in Fig. 12. 
 }
    \label{LSCZ_erosion}
\end{figure}

\begin{figure}
    \centering
    \subfigure[Erosion]{\includegraphics[width=.48\linewidth]{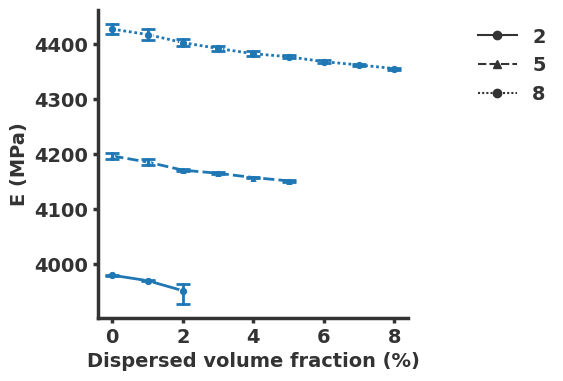}\label{macroE_noIntph_Erosion}}
    \subfigure[Rupture]{\includegraphics[width=.48\linewidth] {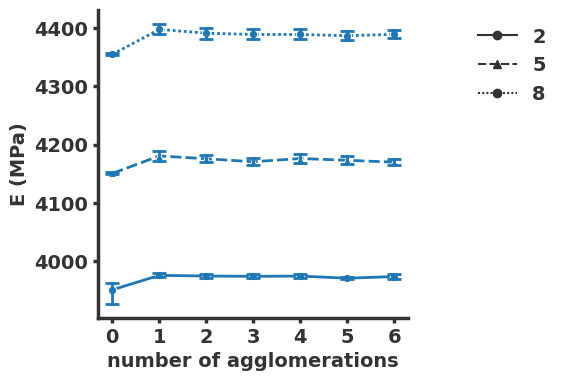}\label{macroE_noIntph_Rupture}}
    \caption{effective Young's modulus for the complete DoE for the simplest 'no Interphase' scheme i.e only two phase material system}
    \label{macroE_noIntph}
\end{figure}

In Figure \ref{LSCZ_rupture_no}, for the simplest no interphase case, in the left most image for a single agglomeration, we see a classical stress concentration zone, similar to that of a single particle under periodic boundary conditions. Going from left to right in Figure \ref{LSCZ_rupture_no}, with increasing levels of rupture, we see different ways in which interaction between stress patterns around individual agglomerations takes place. We observe stress amplification through superposition between stress concentration zones (shades of color orange) around individual agglomeration ($LSCF \geq 1$). We also observe stress shielding/reduction due to stress concentration zones (shades of color orange) originating from one agglomeration superposing with stress relaxation zones (shades of color purple) originating from another agglomeration.  The resultant of these stress pattern interactions depends on specific microstructure and relative positioning and orientation of agglomerations. In Figure \ref{LSCZ_erosion_no}, we observe that with increasing erosion levels and particle dispersion, stress concentration zones and resulting interactions are gradually spread out throughout the domain. There is an associated slight decrease in the effective composite Young's modulus with increasing dispersion, as shown in Figure \ref{macroE_noIntph_Erosion}, suggesting that average microstructure linear response is dominated by stress concentration amplification. Based on Figure \ref{macroE_noIntph_Rupture}, no such conclusion can be drawn in case of rupture. 

Considering realistic composites with graded interphase, we see an expansion in stress concentration zones, *** NOTE COMMENT TO FIX OR DELETE*** and more prominently so around agglomerations (refer Figure \ref{LSCZ_rupture_fast} and \ref{LSCZ_rupture_slow}) than around isolated particles (refer Figure \ref{LSCZ_erosion_fast} and \ref{LSCZ_erosion_slow}). We also notice higher levels of stress concentration values in fast interphase scheme compared to slow interphase scheme, which is more apparent in high erosion levels (compare images in last two columns in Figure \ref{LSCZ_erosion_fast} and \ref{LSCZ_erosion_slow}). **** This behavior can be attributed to the fact that faster $E$ gradients result in narrower interphase regions with steep drop in stiffness values compared to slower gradients.  
While the local stress concentration factors increase over baseline (no-interphase case) as the graded interphases are included (Fig 13), at the same time the addition of the graded interphase zones increase the local moduli by definition (Fig 15). Thus there is a competing effect of the increased local interphase stiffnesses pushing the effective modulus higher while the increased LSCZ would tend to decrease effective modulus. From Figure 7, we observe in the no-interphase case that the LSCZ results in a slow decrease moduli with erosion, while for all interphase cases, the effective moduli increase with erosion, indicating that the interphase effect wins the competition. Additionally Figure 7 confirms that the higher LSCZ contributions observed for the fast-decay interphase (Fig 13) result in a slower relative increase of modulus with erosion.

\begin{figure}
    \centering
    \subfigure[Erosion with slow stiffness gradients]{\includegraphics[width=\linewidth]{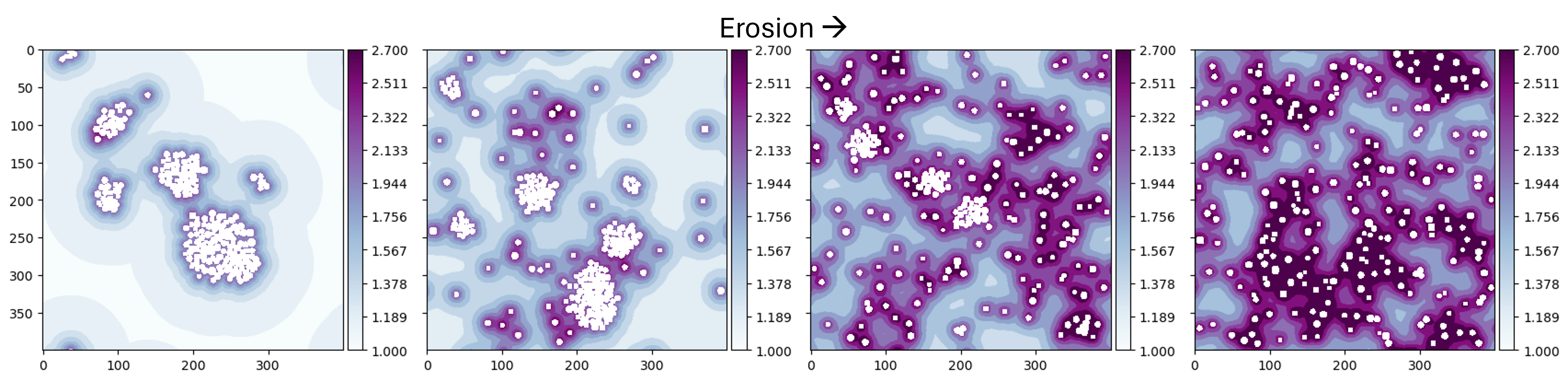}\label{E_Slow_Erosion}} 
    \subfigure[Erosion with fast stiffness gradients]{\includegraphics[width=\linewidth]{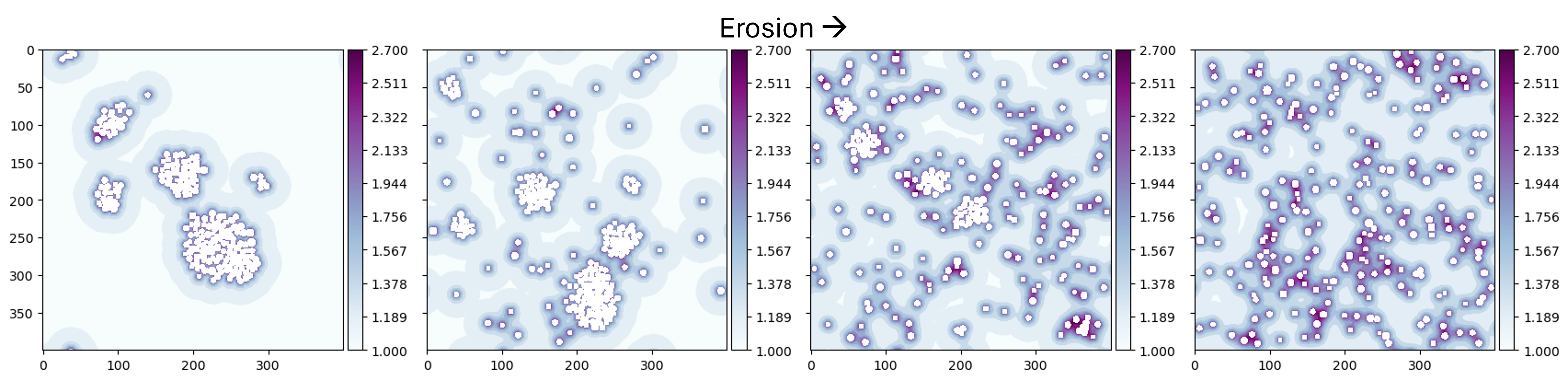}\label{E_Fast_Erosion}} 
    \subfigure[Rupture with slow stiffness gradients]{\includegraphics[width=\linewidth]{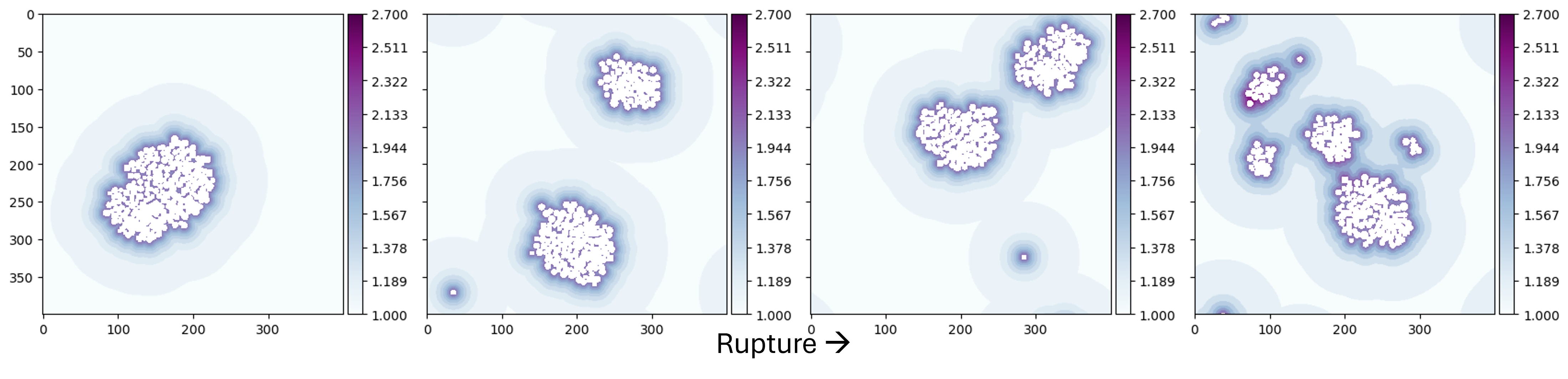}\label{E_Slow_Rupture}} 
    \subfigure[Rupture with fast stiffness gradients]{\includegraphics[width=\linewidth]{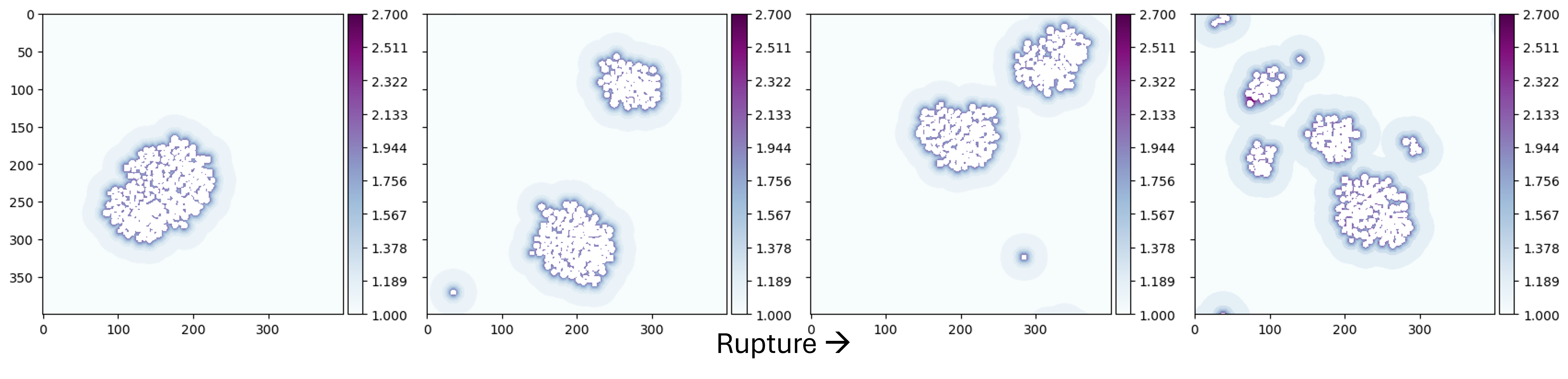}\label{E_Fast_Rupture}} 
    \caption{Stiffness gradients comparison between Erosion and Rupture combined with slow and fast gradients }
    \label{stiffness_gradients}
\end{figure}

\subsection{Yielding Mechanism} \label{yielding_mechanism}
To understand yield behavior, we focus on two parts of the yielding process, local yield initiation and its progression with loading. Zones with stress concentration values higher than 1 have higher probability, compared to those with values lower than 1, to undergo local yielding at any given macro-stress value or applied load conditioned on the local yield strength of the material. However, the local yield limits are determined by the graded yield interphase. There is thus another competition between the local stress concentrations which tend to decrease effective yield strength and the percolation of interphase zones with higher local yield stresses which tend to increase effective yield strength of the composite. The trends observed in effective yield stresses are connected to these two factors as dispersion changes via rupture and erosion.

\begin{figure}
    \begin{longtable}{cc}
            \begin{tabular}{c}
                    \subfigure[$S_{11}= 69.65 MPa $ \text{and} $S_{v-m}= 62.44 MPa$]{
                        \includegraphics[height=0.28\textheight]{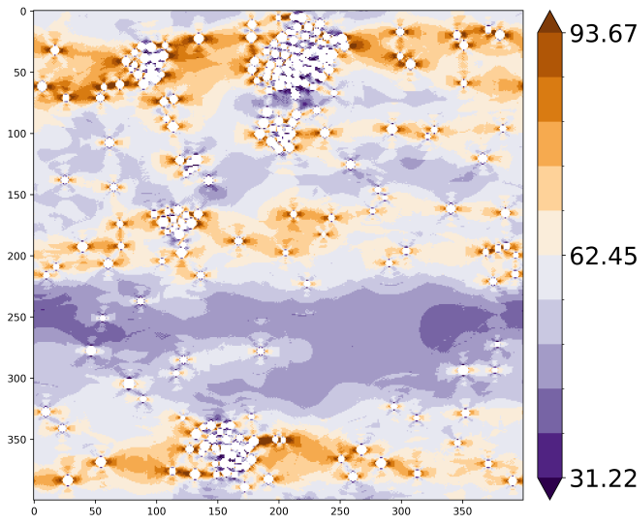}
                        \label{Yield_c5}} \\
                    \subfigure[$S_{11}= 107.99 MPa $ \text{and} $S_{v-m}= 96.81 MPa$]{
                        \includegraphics[height=0.28\textheight]{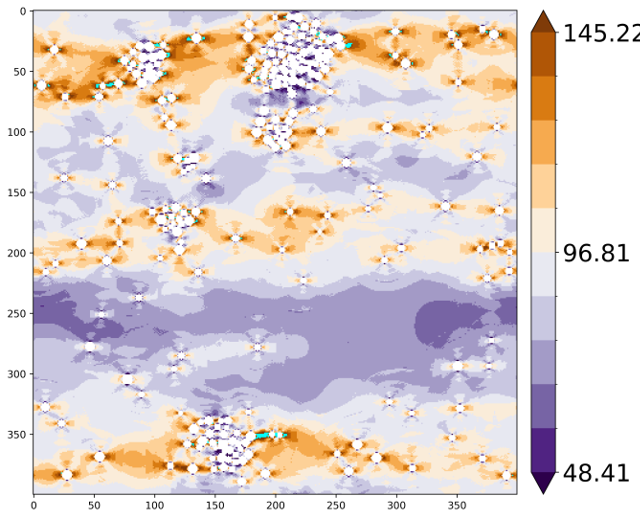}
                        \label{Yield_c6}} \\
                    \subfigure[$S_{11}= 126.39 MPa $ \text{and} $S_{v-m}= 113.28 MPa$]{
                        \includegraphics[height=0.28\textheight]{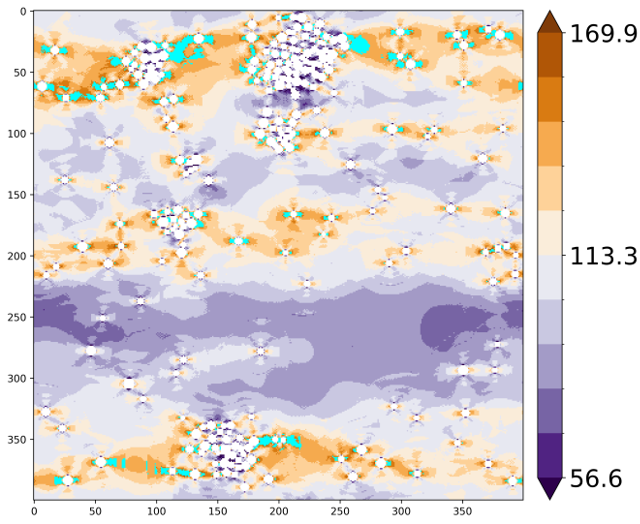}
                        \label{Yield_r16}}
            \end{tabular}
            &
            \begin{tabular}{c}
                    \subfigure[$S_{11}= 142.58 MPa $ \text{and} $S_{v-m}= 127.69 MPa$]{
                        \includegraphics[height=0.28\textheight]{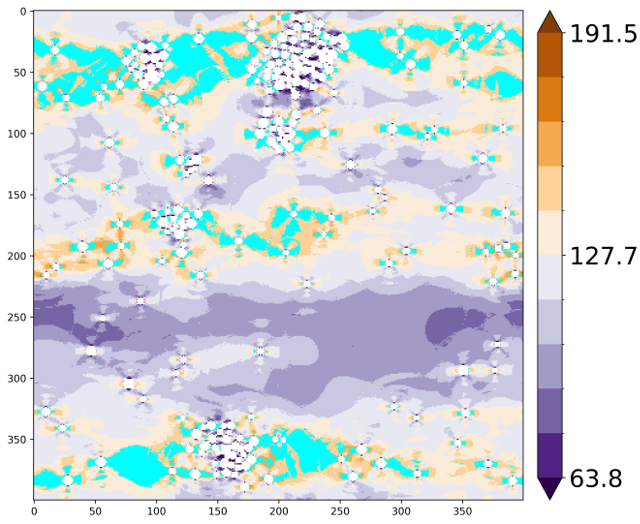}
                        \label{Yield_r19_extra}} \\
                    \subfigure[$S_{11}= 156.96 MPa $ \text{and} $S_{v-m}= 140.32 MPa $]{
                        \includegraphics[height=0.28\textheight]{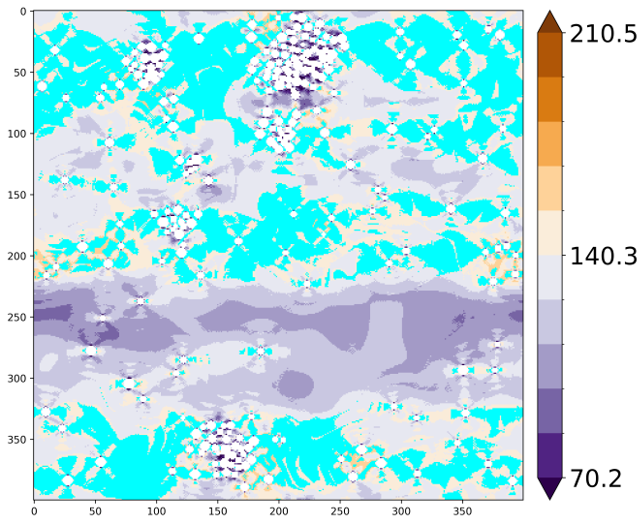}
                        \label{Yield_c7}} \\
                    \subfigure[$S_{11}= 166.82 MPa $ \text{and} $S_{v-m}= 148.70 MPa$]{
                        \includegraphics[height=0.28\textheight]{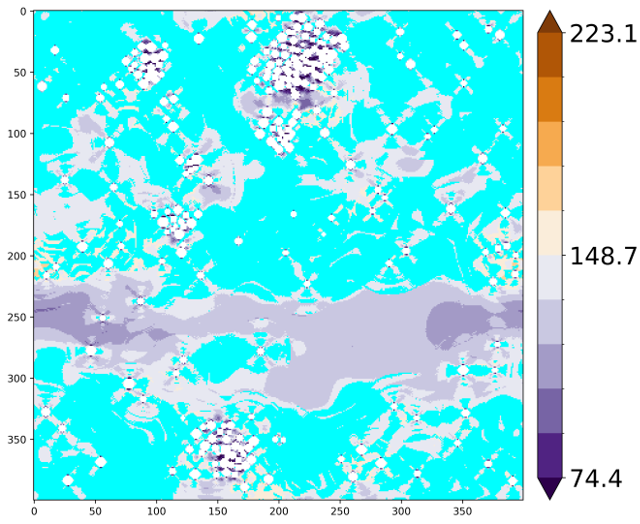}
                        \label{Yield_r17}}
            \end{tabular}
    \end{longtable}
    \caption{Local yielding initiation and progression}
    \label{local_yielding}
\end{figure}

Figures \ref{Yield_c5}, \ref{Yield_c6}, \ref{Yield_r16}, \ref{Yield_r19_extra}, \ref{Yield_c7}, \ref{Yield_r17} show yield initiation progression along with values for axial and von mises macro scale stress values, $\sigma_{11}^{macro}$ nd $\sigma_{v-m}^{macros}$ , in a specific microstructure. It is observed that local yield initiation takes place in the regions of high local stress concentration next to particle aggregates. With increasing loading (figures a to f), these sites of local yield initiation progress through the microstructure to connect and form a yielded region network. 

***CAPTION for Fig 16: Local yielding initiation and progression for TVF $5\%$, DVF of $2\%$ and 6 agglomerations where the macro scale loading direction stress and von mises stress are indicated in each sub-caption. Color bars show the XXXXX local stress. The cyan color in sub-figures show local material that has passed its yield point. ***

\begin{figure}
    \centering
    \subfigure[Linear Stress Concentration]{\includegraphics[width=.48\linewidth]{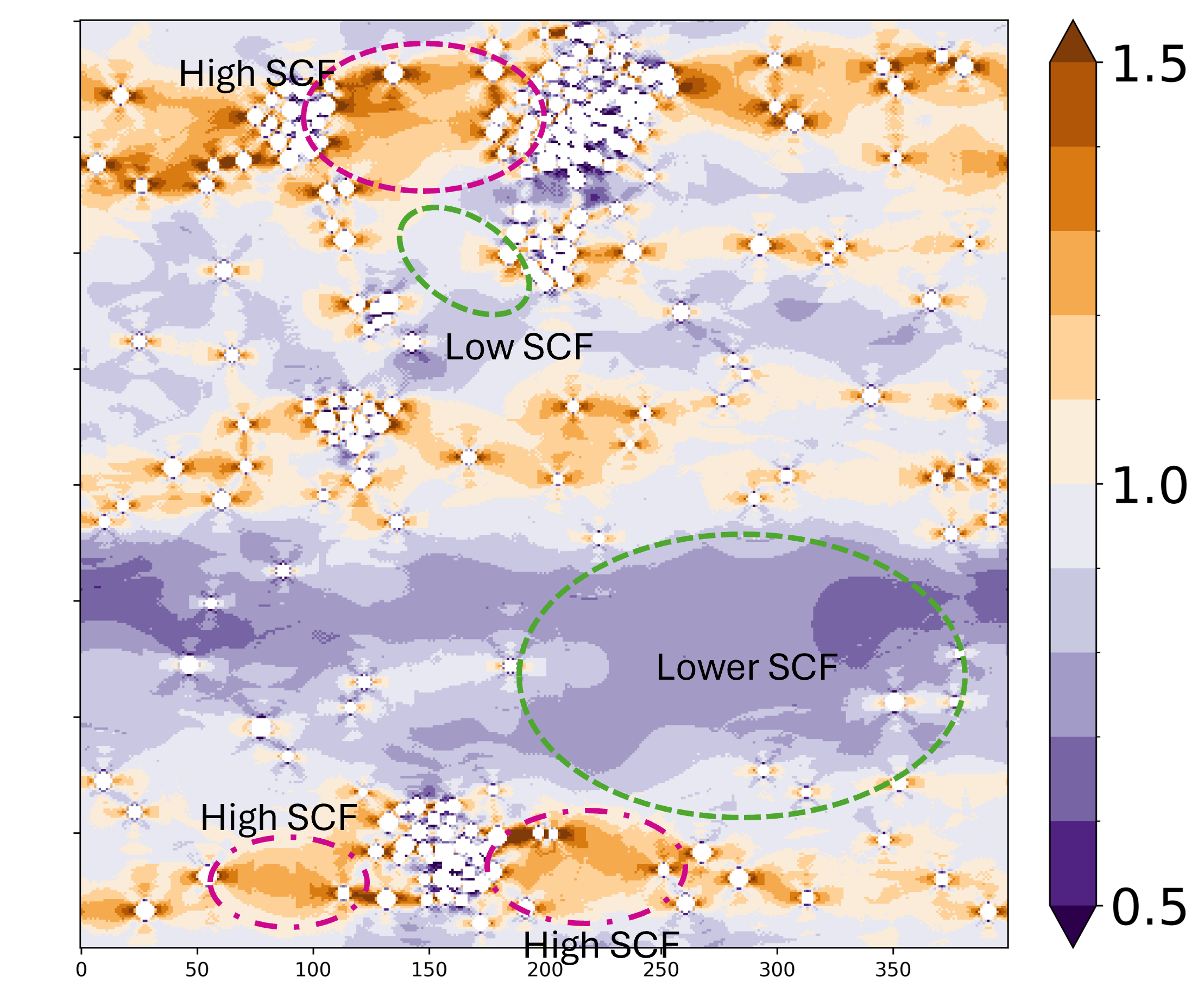}\label{LSCF_for yield}}
     \subfigure[Yield stress interfacial gradients]{\includegraphics[width=.48\linewidth]{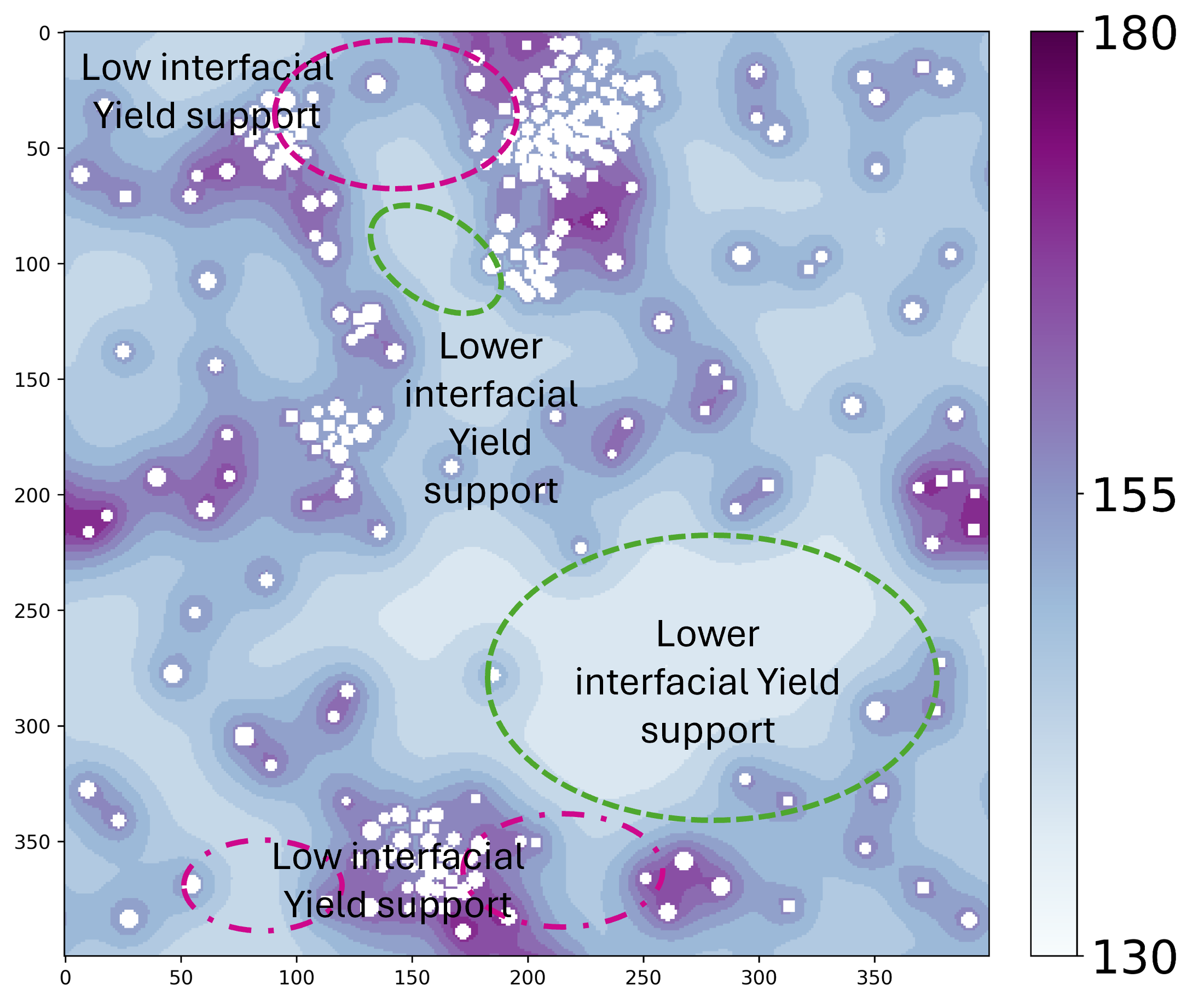}\label{YieldIP_for yield}}
    \caption{(a) Linear stress concentration map for the microstructure from Figure \ref{local_yielding} where regions of higher and lower SCF zones are indicated (b) Yield strength Interphase map where regions of higher and lower interfacial yield values are indicated by the color bar. Regions A and B described in the text.}
    \label{SCF_YieldIP}
\end{figure}

In Figure \ref{SCF_YieldIP}, we observe a complex interplay between local networks of stress concentrations (LSCF) and networks of locally increased interfacial yield stress values (s\_y\_micro). Fig 17b shows explicitly the magnitude of s\_y\_micro and its variability due to the relative locations of the particles in the microstructure.  Local yielding follows a specific path that can be described by a combination of LSCF and $\sigma^{micro}$. To visualize this yielding path we devised a quantity called yield resistance, which describes macroscopic applied stress, $\sigma^{macro}$ , required to cause local yielding for each individual material point(each pixel in this study). 

The relation between local von-mises (v-m) stress $\sigma_{v-m}^{micro}$,  $\sigma^{macro}$ and field distribution of local stress concentration factors (LSCFs) is given by
\begin{equation} \label{LSCF}
\sigma_{v-m}^{micro} = LSCF \times \sigma^{macro}    
\end{equation}
Local yielding takes place if the condition below is satisfied
\begin{equation} \label{yielding_condition}
\sigma_{v-m}^{micro} \geq \sigma_{y}^{micro}    
\end{equation} 

Combining Equation \ref{LSCF} and \ref{yielding_condition} 
\begin{equation} \label{susceptibility}
    \sigma^{macro} \geq \frac{\sigma_{y}^{micro}}{LSCF}
\end{equation}

In Equation \ref{susceptibility}, the right hand side denotes the quantity local resistance to yielding. When the right hand side of Eqn (7) is high, when LCSF is low and sigma\_y\_micro is high (such as area A in Figure 17), the region is very resistant to yielding and is among the last places in the sample to experience yield. If the right hand side of Eqn (7) is low, when LCSF is high and sigma\_y\_micro is low (such as area B in Figure 17), the region has low resistance to yielding and yields early in the progression. Due to the complex microstructural dispersion variability, the trade-off between LCSF and sigma\_y\_micro is not obvious throughout the structure, making the use of Eqn 7 valuable. See also figure XX in the SI, which replots Fig 16 using the right hand side of eqn 7 to visualize the yield resistance maps for this case.

\begin{figure}
    \begin{longtable}{cc}
            \begin{tabular}{c}
                \subfigure[]{
                    \includegraphics[height=0.28\textheight]{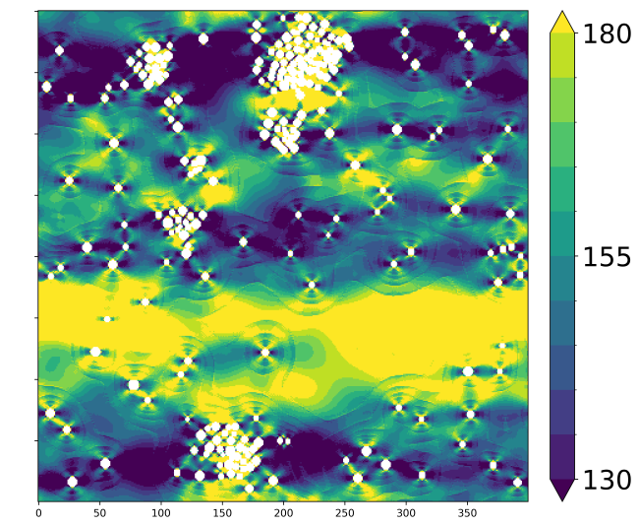}
                    \label{YieldSusceptibility_c5}} \\
                \subfigure[]{
                    \includegraphics[height=0.28\textheight]{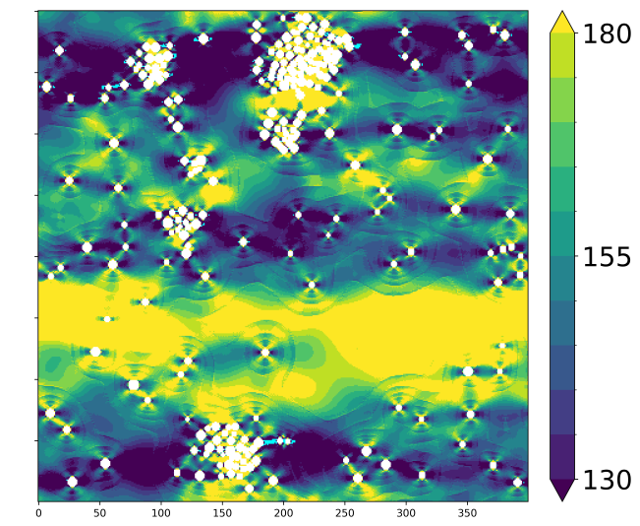}
                    \label{YieldSusceptibility_c6}} \\
                \subfigure[]{
                    \includegraphics[height=0.28\textheight]{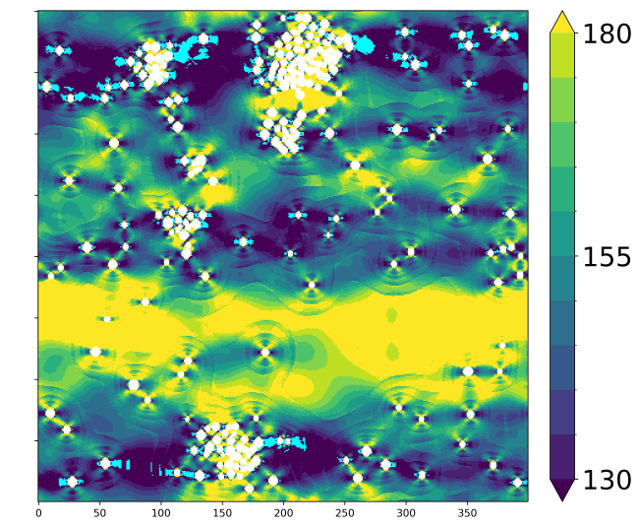}
                    \label{YieldSusceptibility_r16}}
            \end{tabular}
            &
            \begin{tabular}{c}
                \subfigure[$S_{11}= $ \text{and} $S_{v-m}=$]{
                    \includegraphics[height=0.28\textheight]{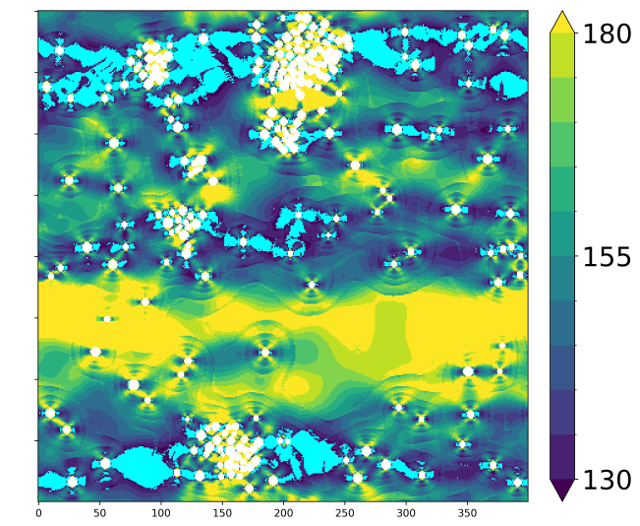}
                    \label{YieldSusceptibility_r19_extra}} \\
                \subfigure[$S_{11}= $ \text{and} $S_{v-m}=$]{
                    \includegraphics[height=0.28\textheight]{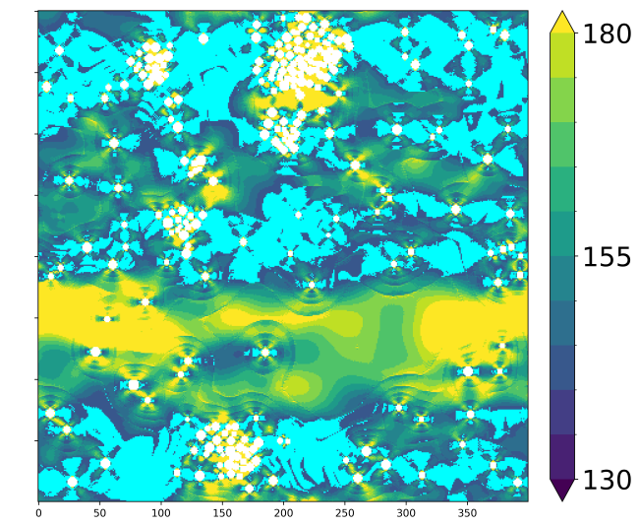}
                    \label{YieldSusceptibility_c7}} \\
                \subfigure[$S_{11}= $ \text{and} $S_{v-m}=$]{
                    \includegraphics[height=0.28\textheight]{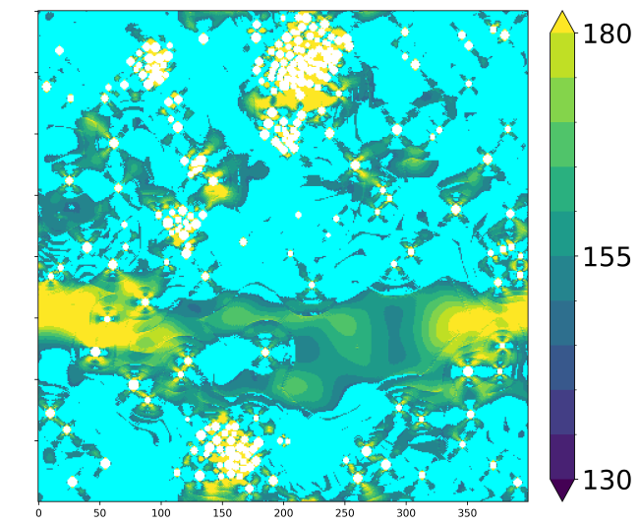}
                    \label{YieldSusceptibility_r17}}
            \end{tabular}
    \end{longtable}
    \caption{Local yielding initiation and progression(Cntd)}
    \label{yield_susceptibility}
\end{figure} 

\subsection{Yield susceptibility maps and yield percolation network}
The field distribution of LSCF within the limit of proportionality depends only on stiffness interfacial gradients $E^{micro}$. Hence LSCZs for the fast and decoupled schemes remain identical until the onset of local yielding. After the initiation of local yielding, the macro stress-strain curve crosses the proportionality limit and enters into the nonlinear deformation zone. Beyond this point, with every load step, load redistribution causes the distribution of LSCF to change. Differences in $\sigma_{y}^{macro}$-VF trends for the two interphase schemes can be attributed to difference in $E$ gradients (see Fig 19 in SI). Hence, we conclude that linear local stress concentrations play a critical role in the yielding mechanism of interfacial regions.

Analysis of yield progression micrographs across DoE data points shows that the yield resistance full field map can act as an indicator of how local yield paths will evolve within the specific microstructure. These maps describe the results from sections \ref{erosionYld} and \ref{ruptureYld}. 

\begin{figure}
    \centering
    \subfigure[Coupled slow]{\includegraphics[width=\linewidth]{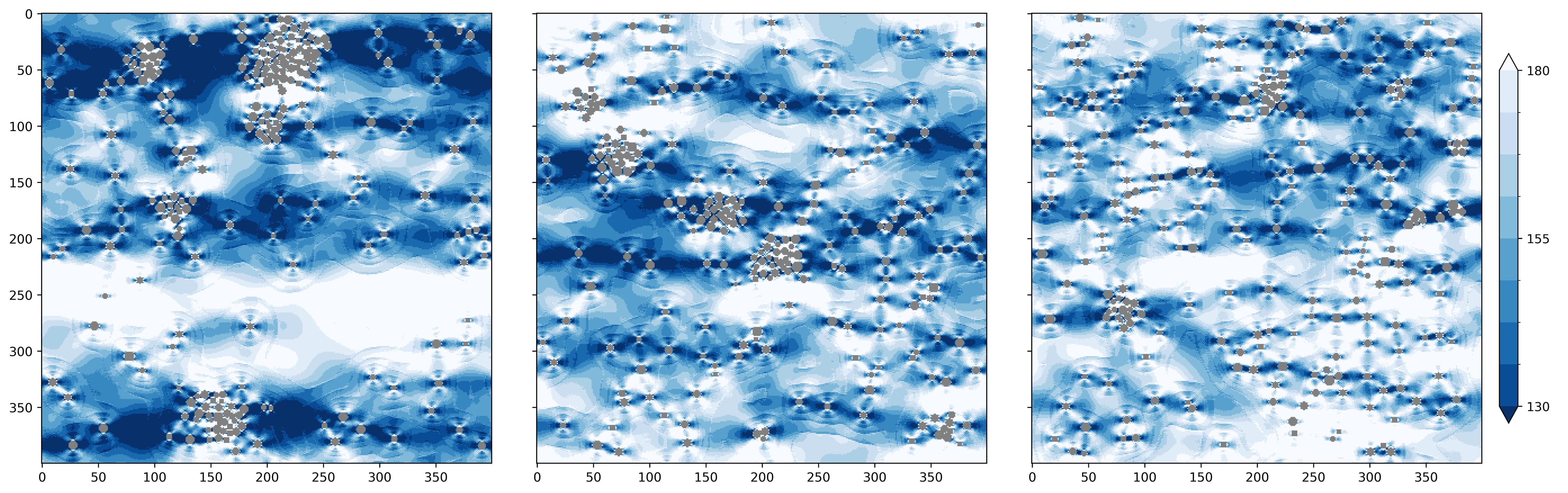}\label{yield_percolation_erosion_slow}}  
    \subfigure[Decoupled]{\includegraphics[width=\linewidth]{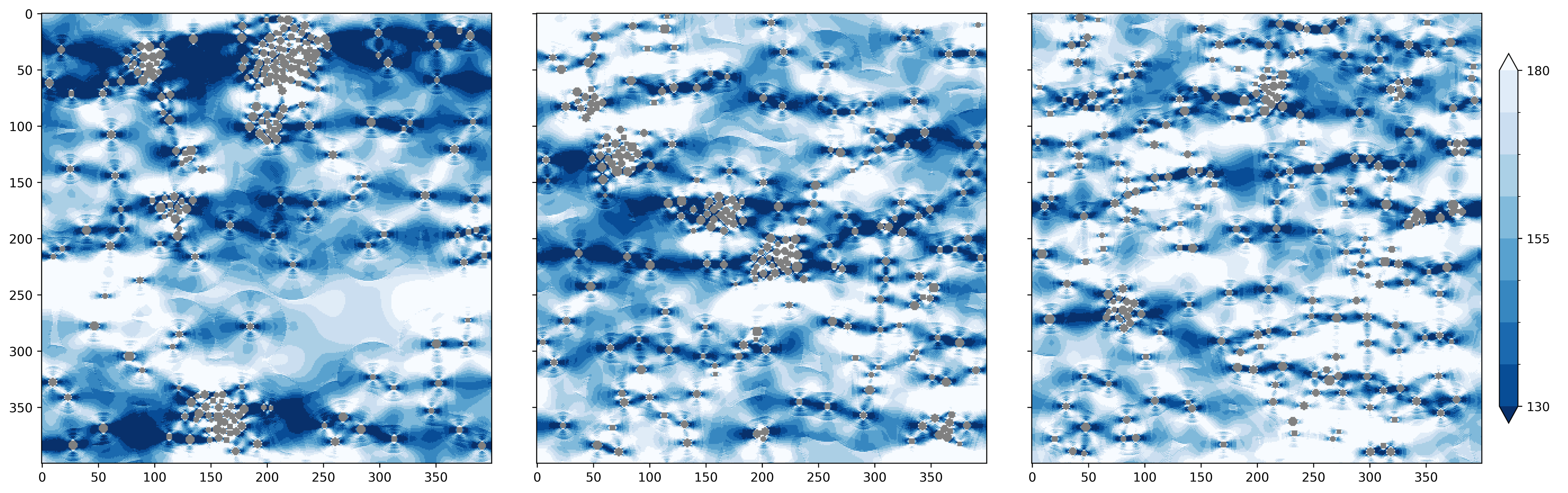}\label{yield_percolation_erosion_decoupled}}  
    \subfigure[Coupled fast]{\includegraphics[width=\linewidth]{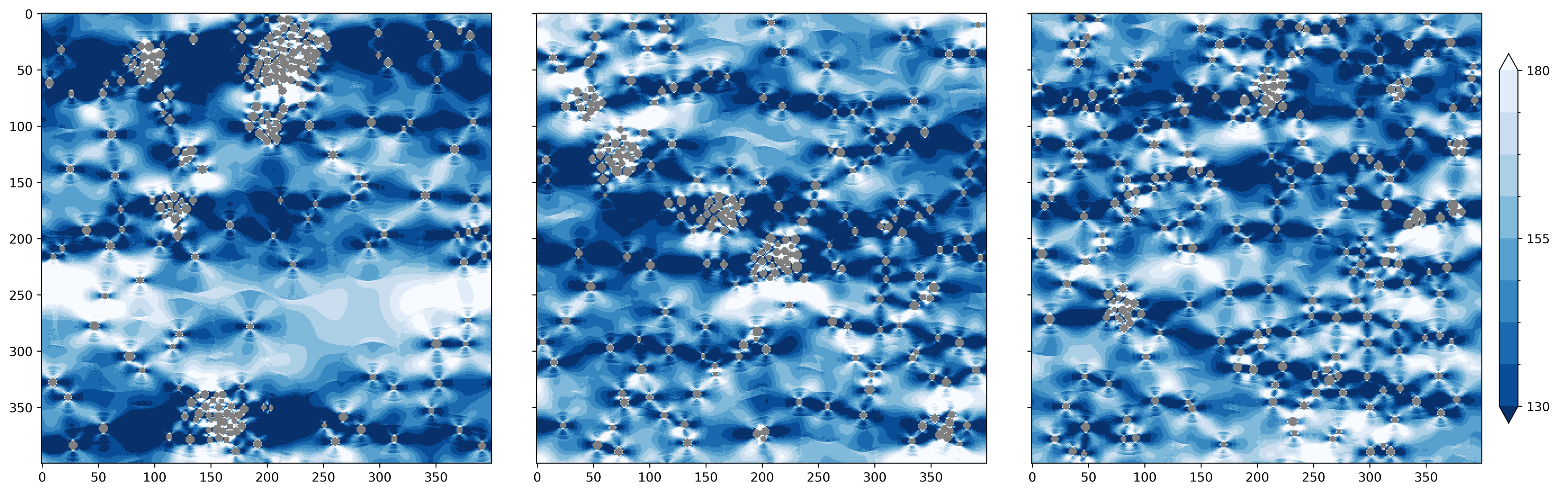}\label{yield_percolation_erosion_fast}} 
    \caption{Yield percolation network map for microstructures obtained by rupture process (going left to right), total VF $5\%$ and (going left to right) DVF $3\%$, $4\%$, $5\%$ and number of agglomerations 6 }
    \label{Yield percolation map}
\end{figure}

\begin{figure}
    \centering
    \includegraphics[width=0.89\linewidth]{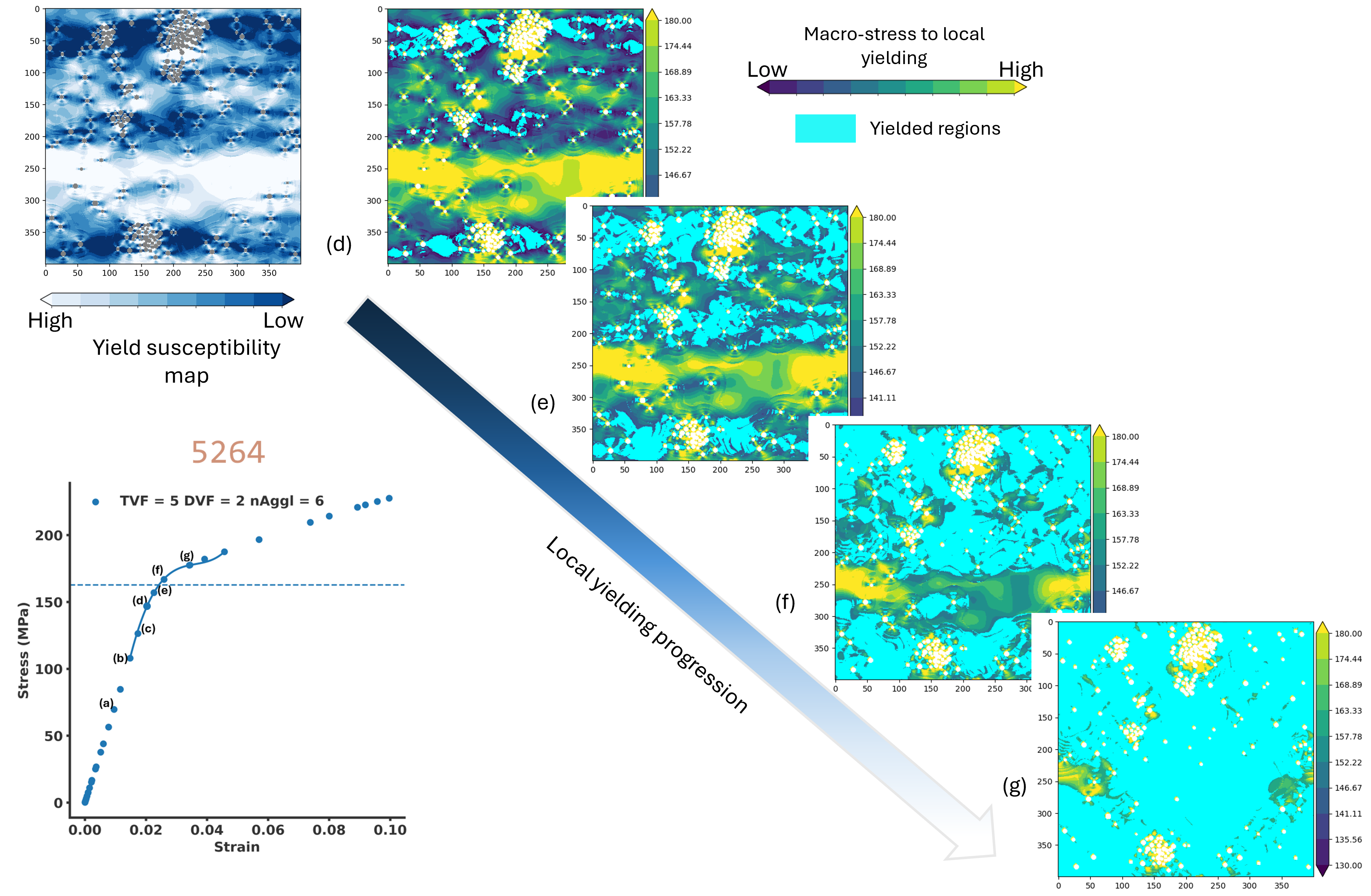}
    \caption{Yield resistance map (a) shoing its power as a blueprint to predict evolution of local yielding shown in subfigures b-e. In b-e, after a pixel has yielded, its color is turned to cyan in these images. (f) shows the macroscopic stress-strain curve of the composite with the location of each subfigure a-e indicated.}
    \label{blueprint}
\end{figure}

Figure \ref{blueprint} shows this relation. Figure \ref{blueprint}(a) shows the resistance map aka blueprint for the yield percolation network. Areas of low resistance, in dark blue, are shown in the yield progression snapshots (Figs 20 b-e) to yield first, while areas of high resistance, in white, are shown to yield last. Thus the yield resistance map allows us to predict existence of the yield percolation network and how it will expand throughout the loading process. This approach also provides a rapid way to compare microstructures with different dispersions to explain the trends observed. 

\section{Conclusion}
This study examines elastoplastic behavior of PNCs across various interphase schemes and dispersion levels, further verifing the critical role of the interphase in reinforcing the mechanical properties of PNCs. The erosion of agglomerated particles facilitates the expanstion and percolation of interphase, thereby augmenting both stiffness and yield more effectively than the rupture process. Rupture exerts a minor effect, impacting the system primarily when most fillers are in an agglomerated state. For stiffness, the the compound effect induced by the interaction of gradient interphase enhances local properties and directly contributes to macro-scale stiffness. Notably, stiffness is found to have a linear correlation with the weighted amount of interphase, suggesting a fast predictive approach for bulk stiffness by simply calculating the weighted amount of interphase. For yield stress, the interphase property enhancements engage in a dynamic competition with local stress zones/concentrations, resulting in intricate bulk yield strength behavior. The results show that while local stress concentrations can lead to a slight decrease of stiffness with erosion of agglomerations, the presence of an attractive interphase combats this decrease; it is noted that interphases with slower decaying gradients have more impact than weaker fast decaying interphases. From a model design perspective, a coupled gradient interphase, which is simpler and faster for computation, can replace a decoupled gradient interphase for stiffness predictions, but not for nonlinear properties due to complex trade-offs. The proposed equivalent uniform interphase provides a reasonable estimation of its corresponding gradient interphase, especially for systems with higher total and dispersed volume fractions, making it an viable alternative for rapid calculations. 

At the same time, the presence of particles and particle aggregates give rise to local stress concentrations. Particle dispersion through the matrix governs how the local field concentrations interact and affect the overall stress distribution. Stress concentrations at the macroscale are widely known to have detrimental effects on properties like fatigue and fracture limits. These behaviors underline the importance of understanding effects from the presence of agglomeration on material system properties.

In these composites, the yield behavior is governed by an intricate competition between local yield thresholds promoted by the interphase and the local stress concentrations caused by varying dispersion conditions. We introduce a concept of yield resistance maps, which can be defined based on the linear regime simulation and takes into account the balance between the local stress concentrations and interphase stiffness and yield values. These maps provide a blueprint from which the evolution of local yielding throughout any given microstructure can be predicted. Overall, this work provides computational evidence that can inform the design of polymer nanocomposites and deepen the understanding of the role and configuration of interfacial gradients. The performance of different interphase schemes can serve as a reference for wet lab experiments when designing PNCs. The methods and results provide guidance to the impact of gradient interphases and their patterns when compared to experimental outcomes.

\section*{Acknowledgements}
The authors would like to gratefully acknowledge the support of AFOSR (Grant No: FA9550-18-1-0381) for this work. The authors also thank Prof. Linda Schadler from the University of Vermont for providing the experimental images of PMMA-\ch{SiO2} nanocomposites. We also thank many insightful discussions with Dr. Richard Sheridan and Dr.Rayehe Karimi Mahabadi from Duke University.

\clearpage
\bibliographystyle{elsarticle-num}
\bibliography{2DAgglElastoplasticIntph.bib}

\clearpage
\section*{Supplementary Figures}

\setcounter{equation}{0}
\setcounter{figure}{0}
\setcounter{table}{0}

\renewcommand{\thefigure}{S\arabic{figure}}

\begin{figure}[htb]
    \centering
    \includegraphics[width=1\linewidth]{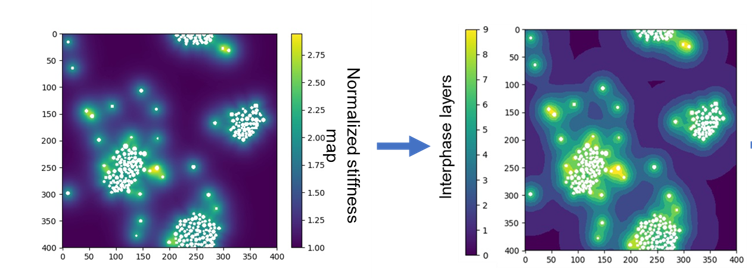} 
    \caption{ An example of property binning from a continuous full field property map. Here, we use 9 bins, with bin number 1 representing material farthest from the nanofillers and with properties closest to that of the matrix and bin number 9 is the portion of interphase closest to nanofillers and displaying the largest change in its local mobility and Tg including the compound effect.}
    \label{SI_binning}
\end{figure}

\begin{figure}[htb]
    \centering
    \includegraphics[width=1\linewidth]{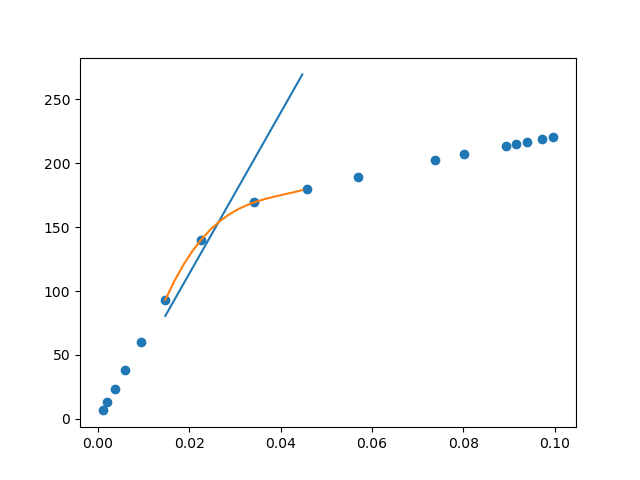} 
    \caption{Illustration of stress-strain data points, yielding region fitting and 0.2\% offset method}
    \label{SIfig_yieldstress}
\end{figure}

Steps of E and yield calculation (Figure \ref{SIfig_yieldstress}): \\
1. Find the linear region. Start from the first point, when (gradient of point n+1/gradient of point n)>1.03, break, linear region ends. \\
2. Calculate E by take average of the gradients of the points in linear region found in step 1.\\
3. Find the yielding region: Start from the first point outside linear region, when (gradient of point n+1/gradient of point n)<1.03, break, fitting region ends.\\
4. Fit the yielding region using 3rd order polynomial.\\
5. Perform a 0.2\% strain offset of the linear region line.\\
6. Find the intersection of the fitted curve in step 4 and offset line in step 5. Yield stress is recorded as the stress value of the intersection.\\

\clearpage
\appendix








\end{document}